\documentclass[preprint,showpacs]{revtex4}

\usepackage{graphicx}
\usepackage{amssymb}
\newcommand{\be}{\begin{eqnarray}}
\newcommand{\ee}{\end{eqnarray}}

\begin{document}

\title{Form factors and transverse charge and magnetization densities in the hard-wall AdS/QCD model}
\author{Chandan Mondal}
\affiliation{Department of Physics, Indian Institute of Technology Kanpur, Kanpur-208016, India.}
%
%\date{\today}
%\maketitle

\begin{abstract}
We present a study of the flavor form factors in the framework of a hard-wall AdS/QCD model and compare with the available experimental data. 
We obtain the flavor form factors by decomposing the Dirac and Pauli form factors for the nucleons using the charge and isospin symmetry. Further, we present a detailed study of the flavor structures of the charge and anomalous magnetization densities in the transverse plane. 
Both the unpolarized and the transversely polarized nucleons are considered here. We compare the AdS/QCD results with two standard phenomenological parametrizations.
%\keywords{charge densities, AdS/QCD, flavor decomposition}
\end{abstract}
\maketitle


%% main text%%%%%%%%%%%%%%%%%%%%%%%%%%%%%%%%%%%%%%%%%%%%%%%%%%%%%%%%%%%%%%
%%%%%%%%%%%%%%%%%%%%%%%%%%%%%%%%%%%%%%%%%%%%%%%%%%%%%%%%%%%%%%%%%%%%%%%%%
%\newpage

\section{ Introduction}

%%%%%%%%%%%%%%%%%%%%%%
In recent years, tremendous interest has grown in AdS/QCD correspondence which has emerged as one of the most promising techniques to investigate the structure of mesons and nucleons.  A weakly coupled gravity theory in $AdS_{d+1}$ can be related to a conformal theory at the $d$-dimensional boundary by the AdS/CFT conjecture \cite{maldacena}. In the last decade, AdS/CFT has been applied to explain several QCD phenomena \cite{PS,costa1}. To apply AdS/CFT in QCD, one needs to break the conformal invariance. 
In the hard-wall model, one sets an IR cutoff at $z_0=1/\Lambda_{QCD}$, while in the soft-wall model, a confining potential in $z$ is introduced to break the conformal invariance.
AdS/QCD gives only the semiclassical approximation of QCD, and it has been developed by  several groups for the baryon 
\cite{BT00,BT01,katz,SS,AC,ads1,ModelII,ads2}. So far, various aspects of hadron properties, e.g., hadron mass spectrum, generalized parton distribution functions (GPDs), meson and nucleon form factors, transverse densities, structure functions, etc., have been  successfully described by AdS/QCD \cite{AC,ads1,ModelII,ads2,AC4,BT1,BT2,deTeramond:2005su,vega,CM,CM2,CM3,HSS,Mondal,Ma_baryon,reso}.
%The first application to nucleon resonances in AdS/QCD framework have been analyzed in \cite{reso}. 
The first computation of nucleon transition form factors in AdS/QCD was described in \cite{BT_reso}.
Recently, a quark-scalar diquark \cite{vega_diquark} and a quark-vector diquark \cite{Maji:2016yqo} models have been developed for the nucleon, where the wave functions are constructed from the soft-wall AdS/QCD correspondence, and these have been extensively used to investigate  many interesting properties of nucleons \cite{CM4,CM5,CM6,CM7}. 
In the meson sector, AdS/QCD has also been successfully applied to predict the branching ratios for decays of $\bar{B^0}$ and $\bar{B_s^0}$ into $\rho$ mesons \cite{ahmady1}, isospin asymmetry and branching ratios for the $B\to K^*\gamma$ decays \cite{ahmady2}, transition form factors \cite{AC2,ahmady3}, etc. The result with the AdS/QCD wave functions agrees very well with the experimental data for $\rho$ meson electroproduction \cite{forshaw}. In the baryon sector, there are many other applications; e.g., semiempirical hadronic transverse momentum density distributions have been calculated in \cite{AC3}, the form factor of spin-$3/2$ baryons ($\Delta$ resonance) and also the transition form factor between $\Delta$ and nucleon have been studied in \cite{hong2}. To study the baryon spectrum at finite temperature, an AdS/QCD model has been proposed in \cite{li}. 
Recently, it has been shown that the superconformal quantum mechanics can be precisely mapped to AdS/QCD \cite{BT_new1}. The superconformal quantum mechanics together with light-front AdS/QCD has revealed the importance of conformal symmetry and its breaking within the algebraic structure for understanding the confinement mechanism of QCD \cite{BT_new2,BT_new3}.

Electromagnetic form factors (EFFs) are the fundamental quantities to understand the internal structure of the nucleons and have been measured in  many experiments. For a detailed review on this subject, we refer the reader to the articles \cite{gao, hyde,perd}. The charge and magnetization densities in the transverse plane are defined as the two-dimensional Fourier transformation of the EFFs. The first moments of the GPDs are related to the EFFs. The transverse densities  are again intimately related to the zero skewness GPDs. Using charge and isospin symmetries, the contributions of individual quarks to the nucleon charge and magnetization densities are obtained from the flavor decompositions of the transverse densities. The densities in the transverse plane corresponding to individual quarks are given by the moment of the GPDs in the transverse impact parameter space \cite{burk}. 
The form factors involve initial and final states with different momenta, thus, three-dimensional Fourier transforms cannot have the interpretation of densities, but at fixed light-front time, the transverse densities have a proper density interpretation \cite{miller09,miller10,venkat}.

The nucleon transverse charge and magnetization densities have been evaluated in \cite{vega} using the holographic model developed in \cite{AC}.
Model-independent charge densities in the transverse plane for nucleons in the infinite-momentum frame have been shown in \cite{miller07}, whereas the transverse charge densities  for a transversely polarized nucleon have been studied in \cite{vande,selyugin}. The long-range behaviors of the unpolarized quark 
transverse charge densities of the nucleons have been investigated in \cite{MVT}. Using methods of dispersion analysis and chiral effective field theory, transverse densities in the nucleon's  chiral periphery [i.e., at a distance $b={\cal{O}}(1/m_\pi)$] have been analyzed in \cite{weiss}. In \cite{silva}, the transverse charge and magnetization densities for the quarks have been studied in a chiral quark-soliton model. Kelly \cite{kelly02} proposed a parametrization of  the nucleon Sachs form factors in terms of  charge and magnetization densities using Laguerre-Gaussian expansion. The flavor dependence of the transverse densities in different models of GPDs has been reported in \cite{liuti, neetika}. 

In this work, we consider a hard-wall AdS/QCD model. Although the hard-wall AdS/QCD model is a simple and useful model to describe various hadronic properties, it has some shortcomings when trying to describe the observed meson spectrum \cite{BT2,deTeramond:2005su}. 
This model for the meson is degenerate with respect to the orbital quantum number $L$, which leads to identical trajectories for pseudoscalar and vector mesons. Thus, it fails to account for the important $L = \vert L^z \vert = 1$ triplet splitting $a$-meson states for different values of $J$. Again, for higher quantum excitations, the spectrum in this model behaves as $M\sim 2n + L$, whereas experimentally the usual Regge dependence is found as $\mathcal{M}^2 \sim n + L$~\cite{Klempt:2007cp}. In the hard-wall model, the radial modes are not well described; therefore, the first radial AdS eigenvalue has a mass $1.77$ GeV, which is very high compared to 
the mass of the observed first  radial excitation of the meson, the $\pi(1300)$. A similar difficulty has also been observed in nucleon resonance where the first AdS radial state has a mass $1.85$ GeV, which is, thus hard to identify with the Roper $N(1440)$ resonance \cite{BT2}. In spite of these shortcomings, many other interesting works have been done for the meson and baryon sectors in the hard-wall AdS/QCD model; see, e.g.,\cite{Wang:2015osq,Zhang:2010bn,Zuo:2009hz,Grigoryan:2007wn,Herzog:2006ra}.

In general, soft-wall AdS/QCD models have some advantages compared to hard-wall models. In particular, the hadronic mass spectrum in soft-wall models 
exhibits Regge-like behavior. There are two different soft-wall AdS/QCD models for the nucleon electromagnetic form factors developed by Brodsky and T\'{e}ramond \cite{BT2} (we refer to them as soft I) and Abidin and Carlson \cite{AC} (we refer to them as soft II). The soft I is developed by weighing the different Fock-state components by the charges/spin projections of the quark constituents as dictated by the SU(6) spin-flavor symmetry. In soft II, the authors have introduced an additional gauge invariant nonminimal coupling term which gives an anomalous contribution to the Dirac form factors. The form of the Pauli form factors in these two models is same but the original difference is in the Dirac form factors. Note that the Pauli form factors in the AdS/QCD models are mainly of phenomenological origin. A study of the flavor decompositions of the nucleon form factors in the soft-wall AdS/QCD models has been done in \cite{CM2,chandan_few}. One can find that the flavor form factors in the soft I are in good agreement with the experimental data, but the soft II deviates from the data of flavor form factors. Only for $F_1^d$ at higher $Q^2$, the soft I deviates from the data, and the soft II gives a better overall description. Again, it can be noticed that the soft II describes well the experimental data of the electric Sachs form factor for the neutron \cite{AC,chandan_few}. But, except for $G_E^n(Q^2)$, the obtained nucleon form factors in the hard-wall model are in better agreement with data compared to the soft II as can be seen in \cite{AC}. Recently, a comprehensive analysis of the nucleon electromagnetic form factors and their flavor decomposition within the framework of light-front holographic QCD including the higher Fock components has been presented in \cite{Sufian:2016hwn}.
In the present work, we compute the flavor decompositions of the nucleon EFFs in the hard-wall AdS/QCD model formulated by Abidin and Carlson \cite{AC} and compare with the experimental data as well as with the soft-wall AdS/QCD models (soft I and soft II). We observe that the individual flavor form factors deviate at higher $Q^2$ from the experimental data, but this model produces desirable data for some ratios of flavor form factors such as $F_1^d/F_1^u$ and $G^d_E/G^d_M$. Except for $F_1^d$, other flavor form factors described by soft I are better than the consequences obtained in the hard-wall model. Then we show a detailed analysis of the flavor-dependent transverse densities and the flavors contributions to the nucleon densities calculated in this model and compare with the two global parametrizations  of Kelly \cite{kelly04} and Bradford $et~ al$ \cite{brad}. It is found that the hard-wall model is able to generate good data for the neutron transverse charge densities which are better than those obtained in the soft-wall models \cite{CM3}.  Using the charge and isospin symmetry, we evaluate the flavor form factors $F_1^q$ and $F_2^q$ for the quarks by decomposing the nucleon form factors $F_1$ and $F_2$ . The Fourier transforms of these EFFs provide the charge and magnetization densities in the transverse plane.

The paper is organized as follows. A brief description of the EFFs in the hard-wall AdS/QCD model has been has provided in Sec.\ref{ads}. The results of the flavor form factors are compared with experimental data in this section. The charge and magnetization densities in the transverse plane for both unpolarized and transversely polarized nucleons have been analyzed in Sec.\ref{density}. We also study the individual flavor contributions in this section. At the end, we give a brief summary in Sec.\ref{concl}.

%%%%%%%%%%%%%%%%%%%%%%
\vskip0.2in
\noindent
%\section{Nucleon and flavor form factors in AdS/QCD}\label{ads}
%%%%%%%%%%%%%%%%%%%%%%%%%%%%%%%%%%%%%%%%%%%%%%%%%%%%%%%%%%%%%%%%
\section{Hard-wall AdS/QCD model for nucleon form factors}\label{ads}
%%%%%%%%%%%%%%%%%%%%%%%%%%%%%%%%%%%%%%%%%%%%%%%%%%%%%%%%%%%%%%%%
For the nucleon electromagnetic form factors, we consider the hard-wall model of AdS/QCD proposed by Abidin and Carlson \cite{AC}. A sharp cutoff in $z$ is introduced in this model, which  breaks the conformal invariance and allows QCD mass scale and confinement. The action in the hard-wall model is written as \cite{AC}
\be
S=\int d^4x dz \sqrt{g}\Big( \frac{i}{2}\bar\Psi e^M_A\Gamma^AD_M\Psi -\frac{i}{2}(D_M\bar{\Psi})e^M_A\Gamma^A\Psi
-M\bar{\Psi}\Psi\Big),\label{action}
\ee
where $e^M_A=z\delta^M_A$ is the inverse vielbein. The covariant derivative is $D_M=\partial_M+\frac{1}{8}\omega_{MAB}[\Gamma^A,\Gamma^B]-iV_M$ where the spin connections are $\omega_{\mu z\nu}=-\omega_{\mu\nu z}=\frac{1}{z}\eta_{\mu\nu}$. The Dirac gamma matrices satisfy the anticommutation relation $\{\Gamma^A,\Gamma^B\}=2\eta^{AB}$. In $d=4$ dimensions, $\Gamma_A=(\gamma_\mu, -i\gamma_5)$. The relevant term in the action in Eq.(\ref{action}) which generates the Dirac form factor $F_1$ is given by
\be
S_D=\int d^4x dz \sqrt{g}\bar\Psi e^M_A\Gamma^AV_M\Psi \label{action_2}.
\ee
However, the action in Eq.(\ref{action}) is unable to produce the spin flip (Pauli) form factors $F_2$. To get the Pauli form factors, one needs to add the following extra gauge invariant term to the action \cite{AC}
\be
\eta_{S,V}\int d^4x dz \sqrt{g}~\frac{i}{2}\bar\Psi e^M_A e^N_B~[\Gamma^A,\Gamma^B]~F_{MN}^{(S,V)}\Psi,\label{action_3}
\ee
where $F_{MN}=\partial_MV_N-\partial_NV_M$, and the isoscalar and isovector components of the vector field are denoted by the indices $S$ and $V$.
This additional term Eq.(\ref{action_3}) to the action also contributes to the Dirac form factors. Thus, the form factors for the proton and neutron in this model are given by \cite{AC}
 \be
 F_1^p(Q^2) &=& C_1(Q^2)+\eta_p C_2(Q^2),\label{F1pM2}\\
 F_1^n(Q^2)&=& \eta_n C_2(Q^2),\label{F1nM2}\\
 F_2^{p/n}(Q^2)&=& \eta_{p/n} C_3(Q^2),\label{F2pM2}
 %F_2^n(Q^2)&=& \eta_n C_3(Q^2),\label{F2nM2}
 \ee
where $\eta_{p/n} C_2(Q^2)$ is the anomalous contribution to the Dirac form factor coming from the additional term to the AdS action in Eq.(\ref{action_3}). The functions $C_i(Q^2)$ are defined as
 \be
 C_1(Q^2)&=&\int dz~\frac{V(Q^2,z)}{2z^3}(\psi_L^2(z)+\psi_R^2(z)),\\
 C_2(Q^2)&=&\int dz~ \frac{{\partial}_z V(Q^2,z)}{2z^2}(\psi_L^2(z)-\psi_R^2(z)),\\
 C_3(Q^2)&=&\int dz~ \frac{2m_nV(Q^2,z)}{z^2}\psi_L(z)\psi_R(z),
 \ee
and the parameters are $\eta_{p}=(\eta_V+\eta_S)/2$, and $\eta_{n}=(\eta_V-\eta_S)/2$. The limit of the integration is zero to the cutoff value $z_0=(0.245~\rm{GeV})^{-1}$. The upper cutoff was fixed in Ref.\cite{AC} to determine the nucleon and rho-meson masses. 
%%%%%%%%%%%%%%%%%%%%%%%%%%%%%%%%%%%%%%%%%%%%%%%%%%%%%%%%%%%
\begin{figure*}[htbp]
\begin{minipage}[c]{0.98\textwidth}
\small{(a)}
\includegraphics[width=7.3cm,height=5.8cm,clip]{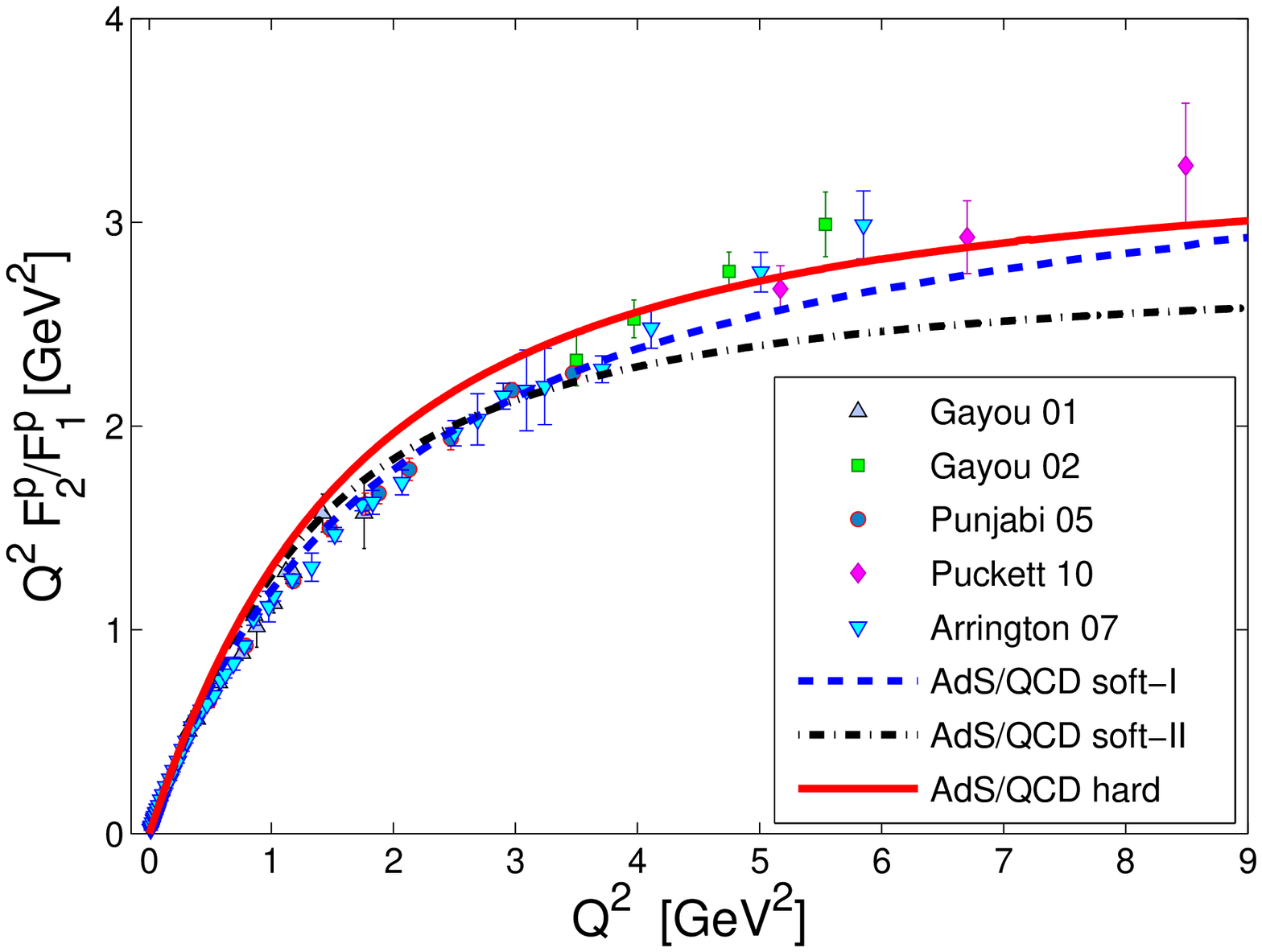}
\hspace{0.1cm}%
\small{(b)}\includegraphics[width=7.3cm,height=5.8cm,clip]{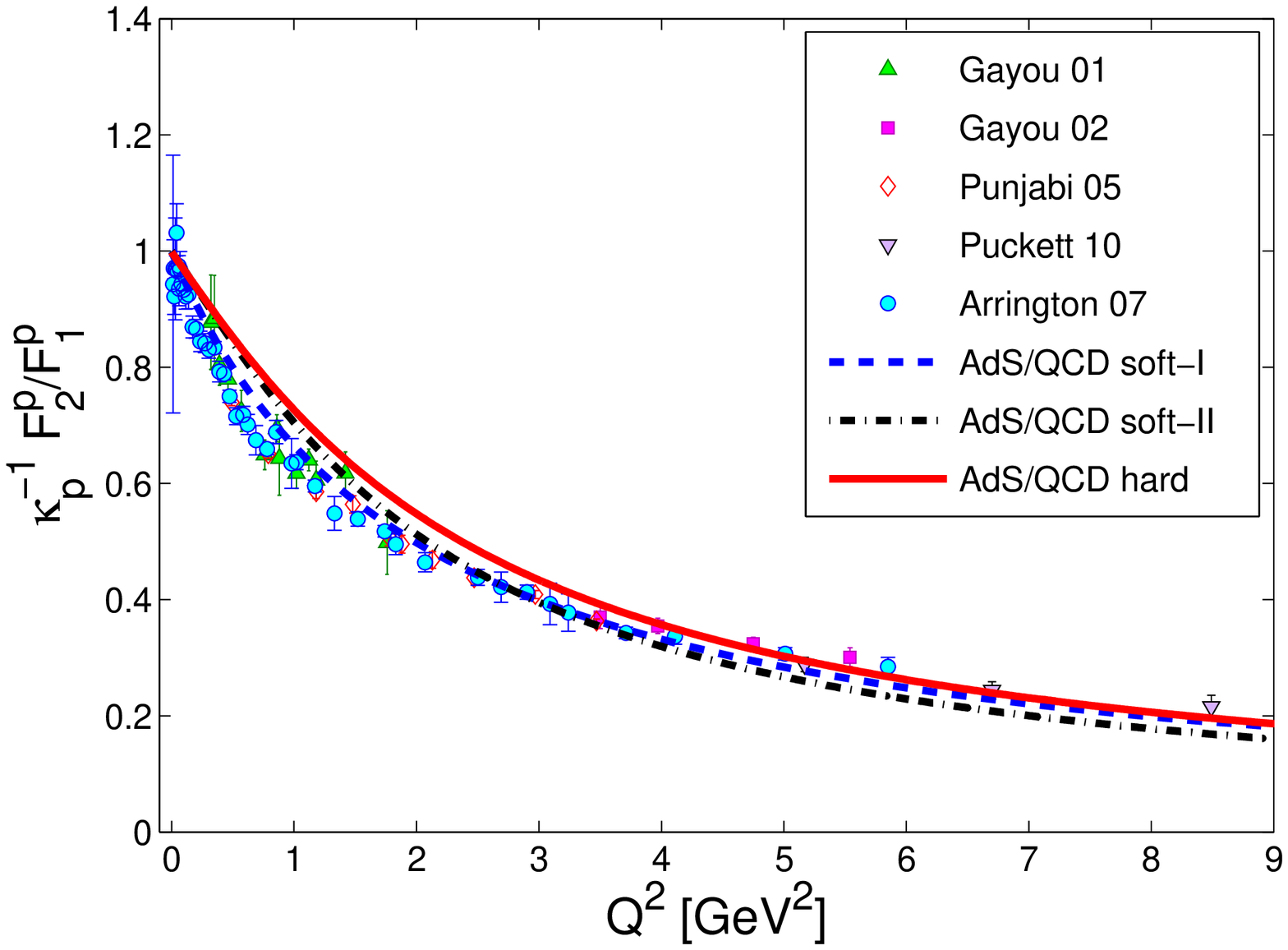}
\end{minipage}
\caption{\label{proton_fit}
The plots show a comparison of the ratio of Pauli and Dirac form factors for the proton between the hard- and soft-wall models in AdS/QCD. (a) The ratio is multiplied by $Q^2=-q^2=-t$ and (b) the ratio is divided by $\kappa_p$. The experimental data are taken from Refs. \cite{Gay1,Gay2,Arr,Pun,Puck}. The solid red line represents the hard-wall AdS/QCD model, the blue dashed line represents the soft-wall AdS/QCD model (soft I) \cite{CM2}, and black dashed-dot line is for the soft-wall model (soft II) \cite{AC}. 
}
\end{figure*}
%%%%%%%%%%%%%%%%%%%%%%%%%%%%%%%%%%%%%%%%%%%%%%%%%%%%%%%%%%%%%%%%%%%%
%%%%%%%%%%%%%%%%%%%%%%%%%%...flavors..%%%%%%%%%%%%
\begin{figure*}[htbp]
\begin{minipage}[c]{0.98\textwidth}
\small{(a)}
\includegraphics[width=7.3cm,height=5.8cm,clip]{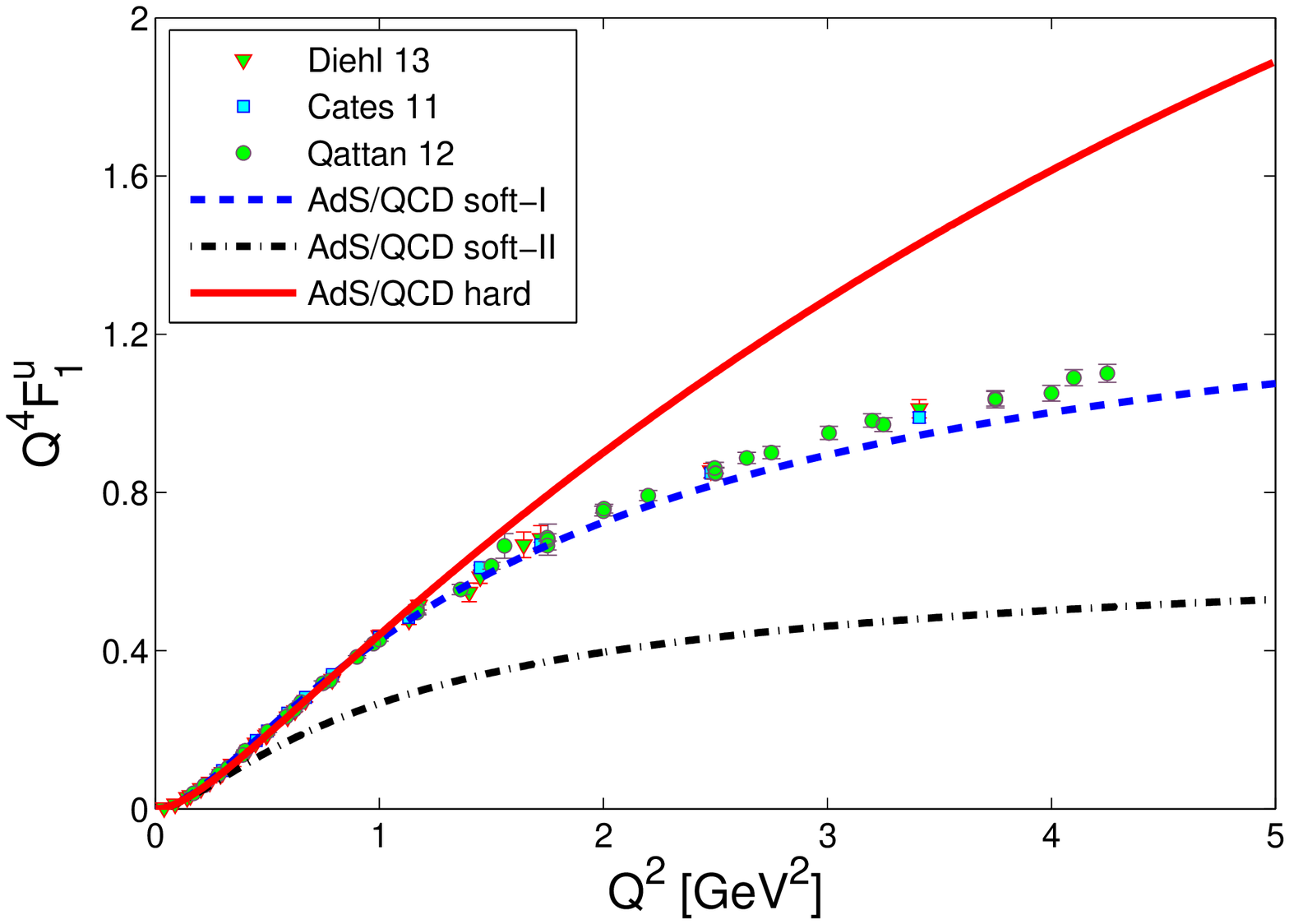}
\hspace{0.1cm}%
\small{(b)}\includegraphics[width=7.3cm,height=5.8cm,clip]{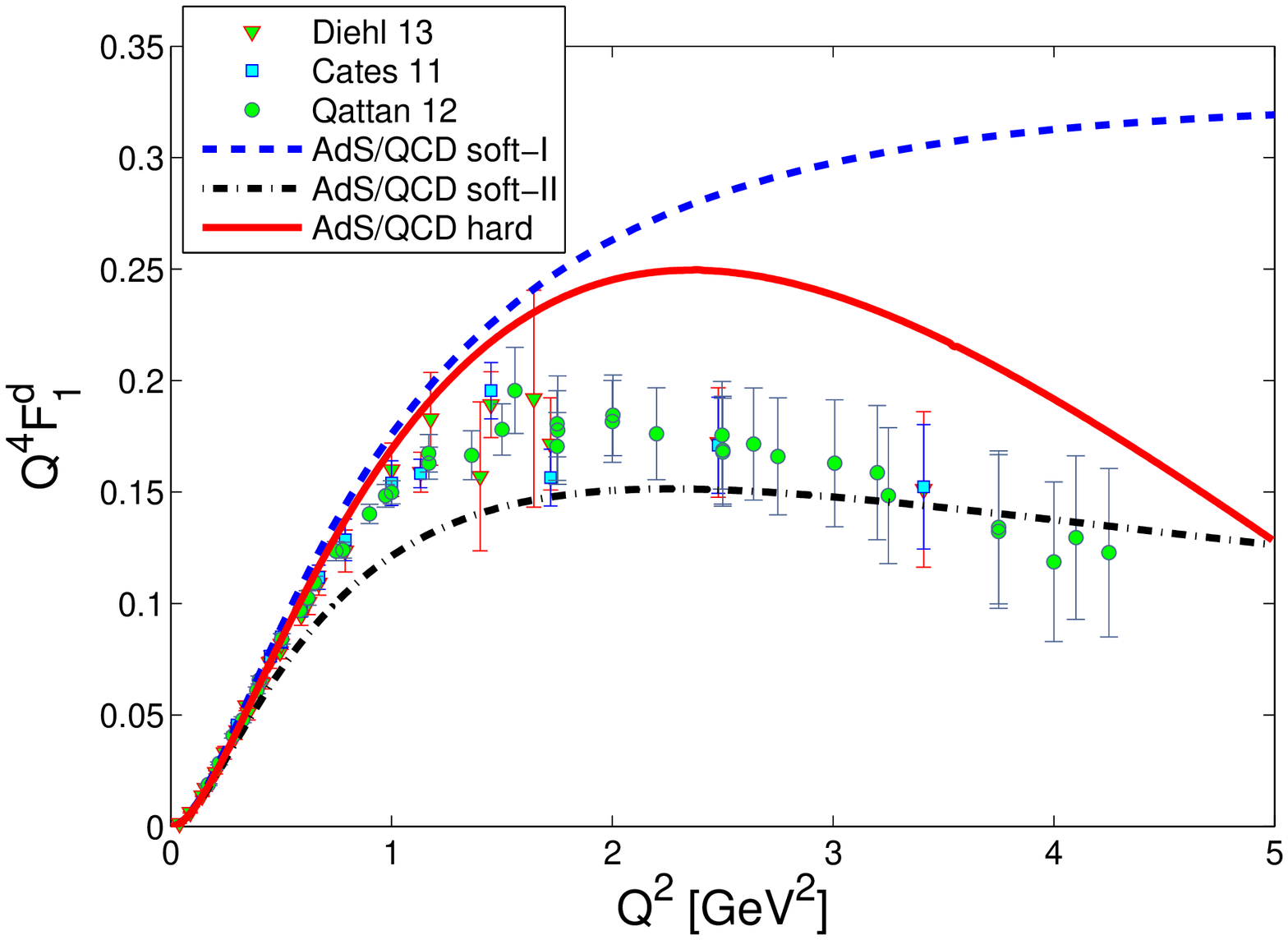}
\end{minipage}
\begin{minipage}[c]{0.98\textwidth}
\small{(c)}
\includegraphics[width=7.3cm,height=5.8cm,clip]{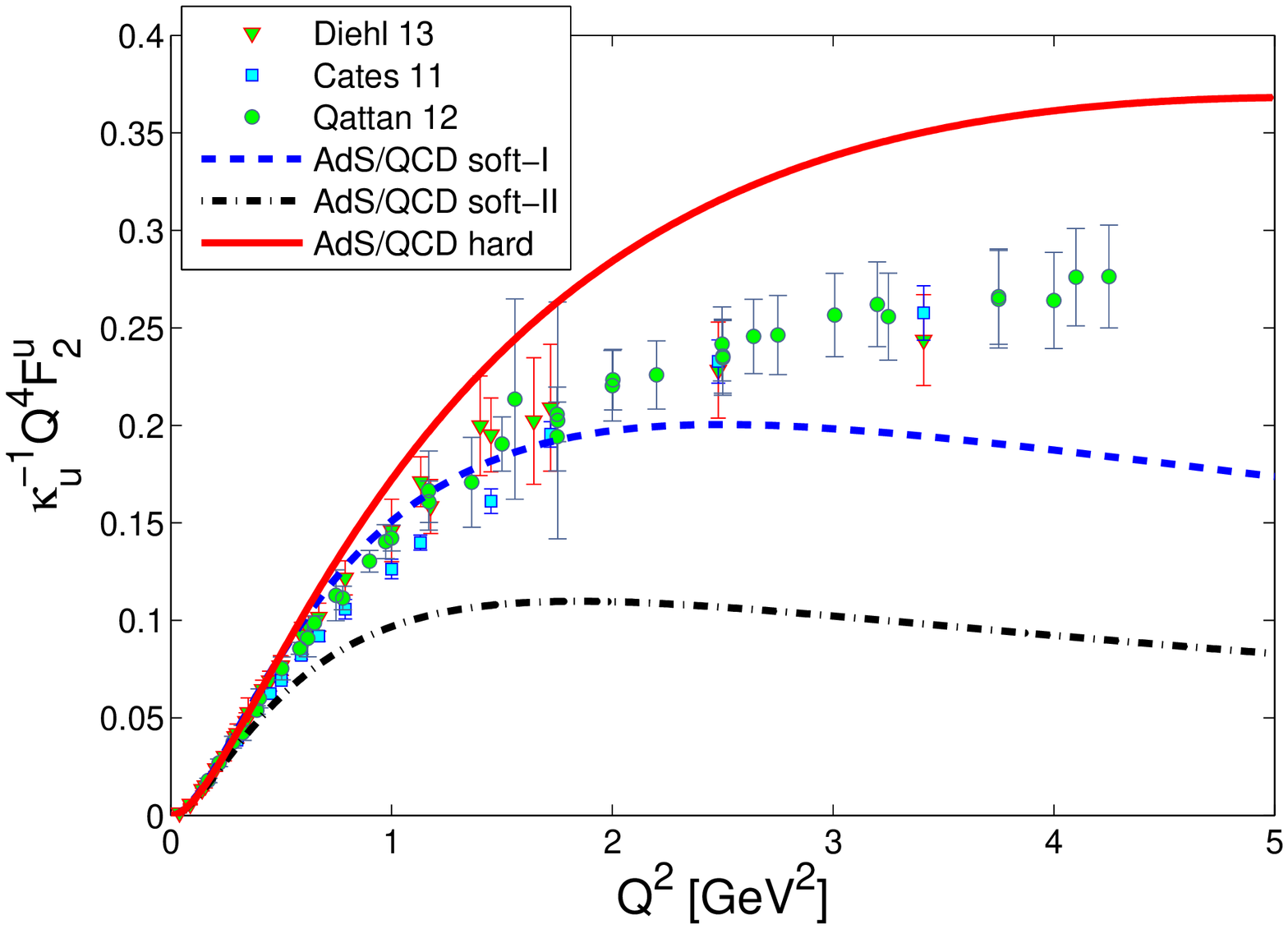}
\hspace{0.1cm}%
\small{(d)}\includegraphics[width=7.3cm,height=5.8cm,clip]{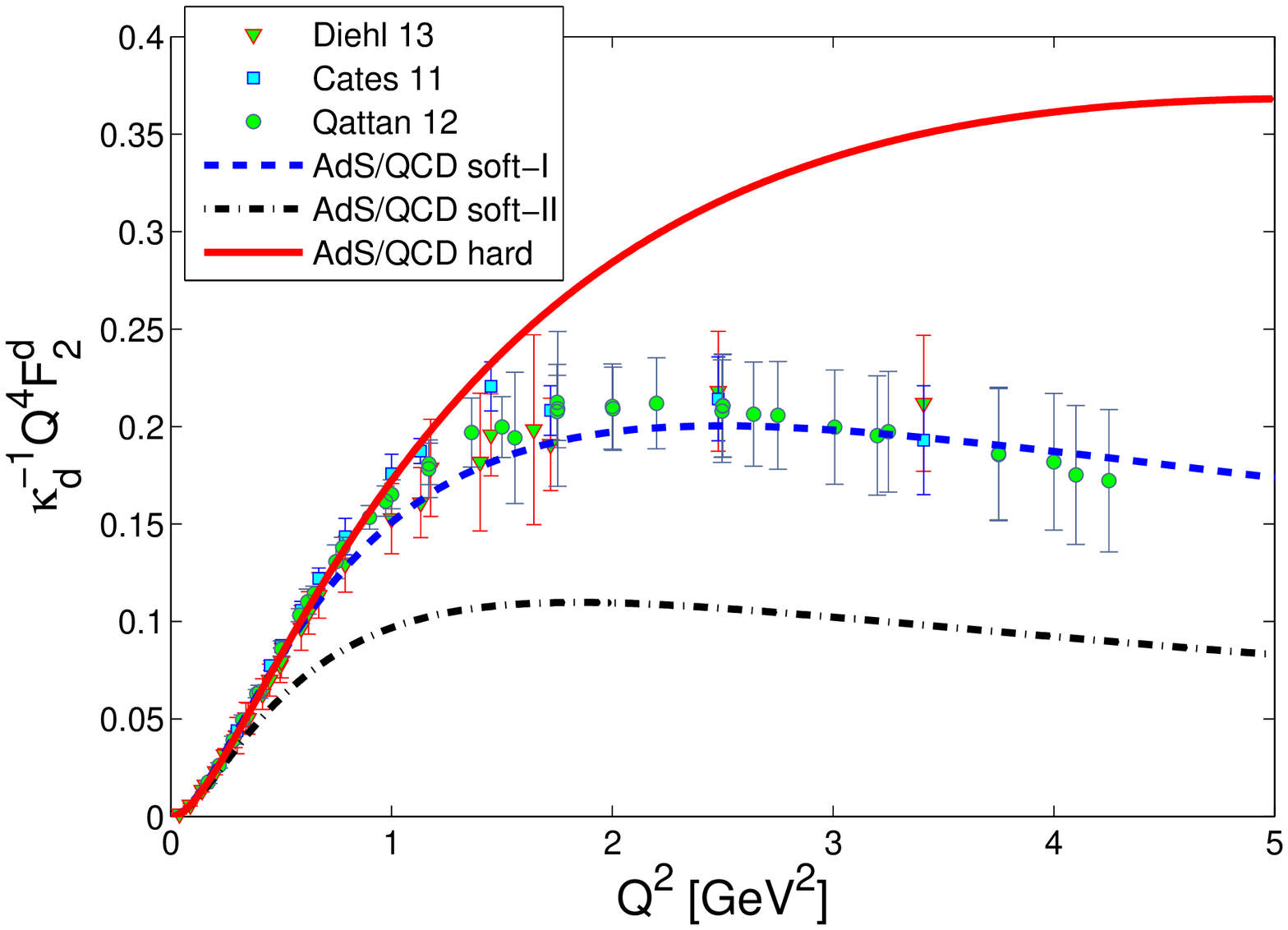}
\end{minipage}
\caption{\label{FF_flavors} Plots of flavor-dependent form factors for the $u$ and $d$ quarks. The experimental data are taken from \cite{Cates,Qattan,diehl13}. The solid red line represents the hard-wall AdS/QCD model, and the blue dashed and black dashed-dot lines represent the soft-wall AdS/QCD models \cite{CM2} and \cite{AC}, respectively.}
\end{figure*} 
%%%%%%%%%%%%%%%%%%%%%%%%%%%%%%%%%%%%%%%%%%%%%%%%%%%%%%%%%%%%%%%%%%%%%%%%%%
\begin{figure*}[htbp]
\small{(a)}\includegraphics[width=7.3cm,height=5.8cm,clip]{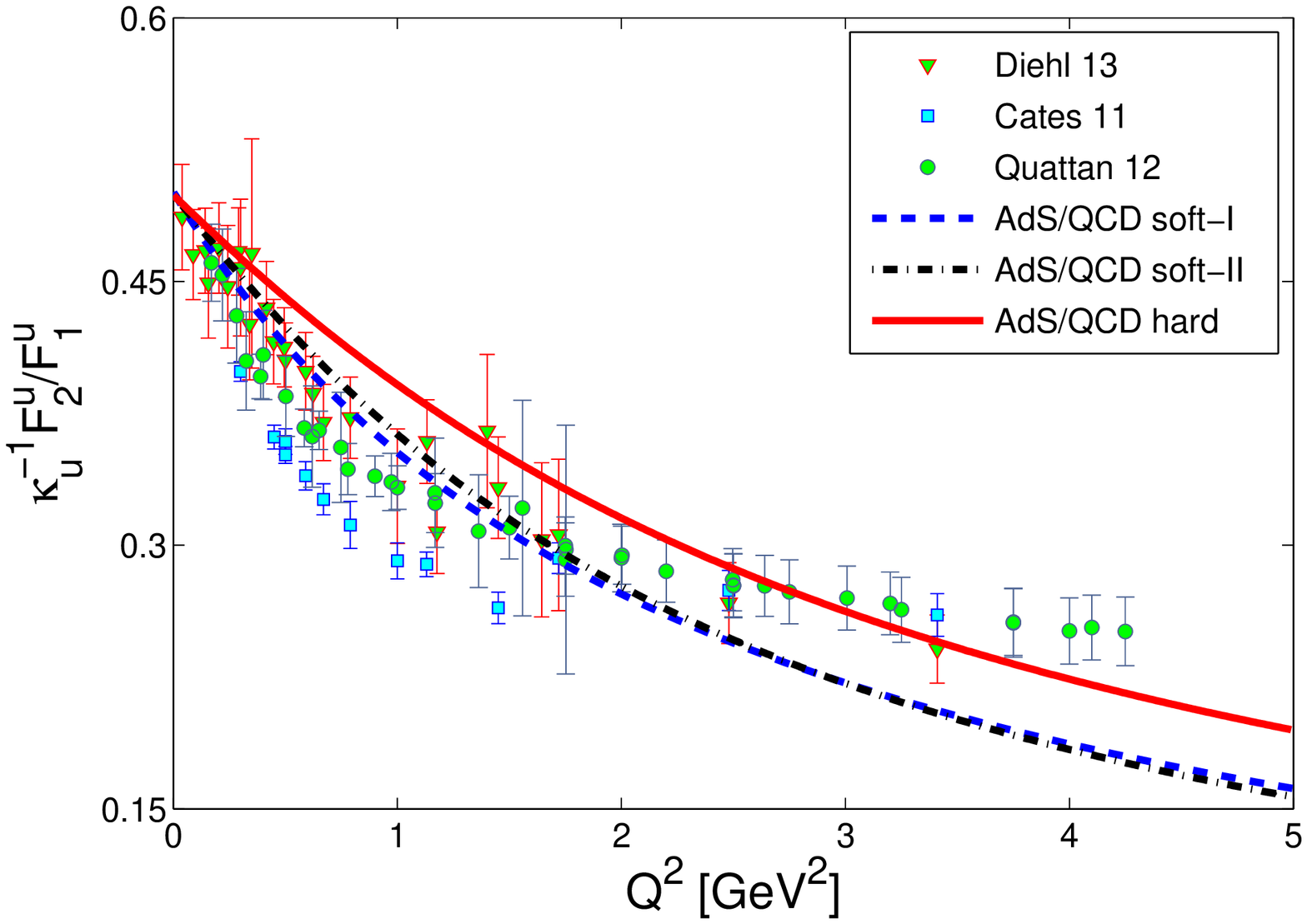}
\small{(b)}\includegraphics[width=7.3cm,height=5.8cm, clip]{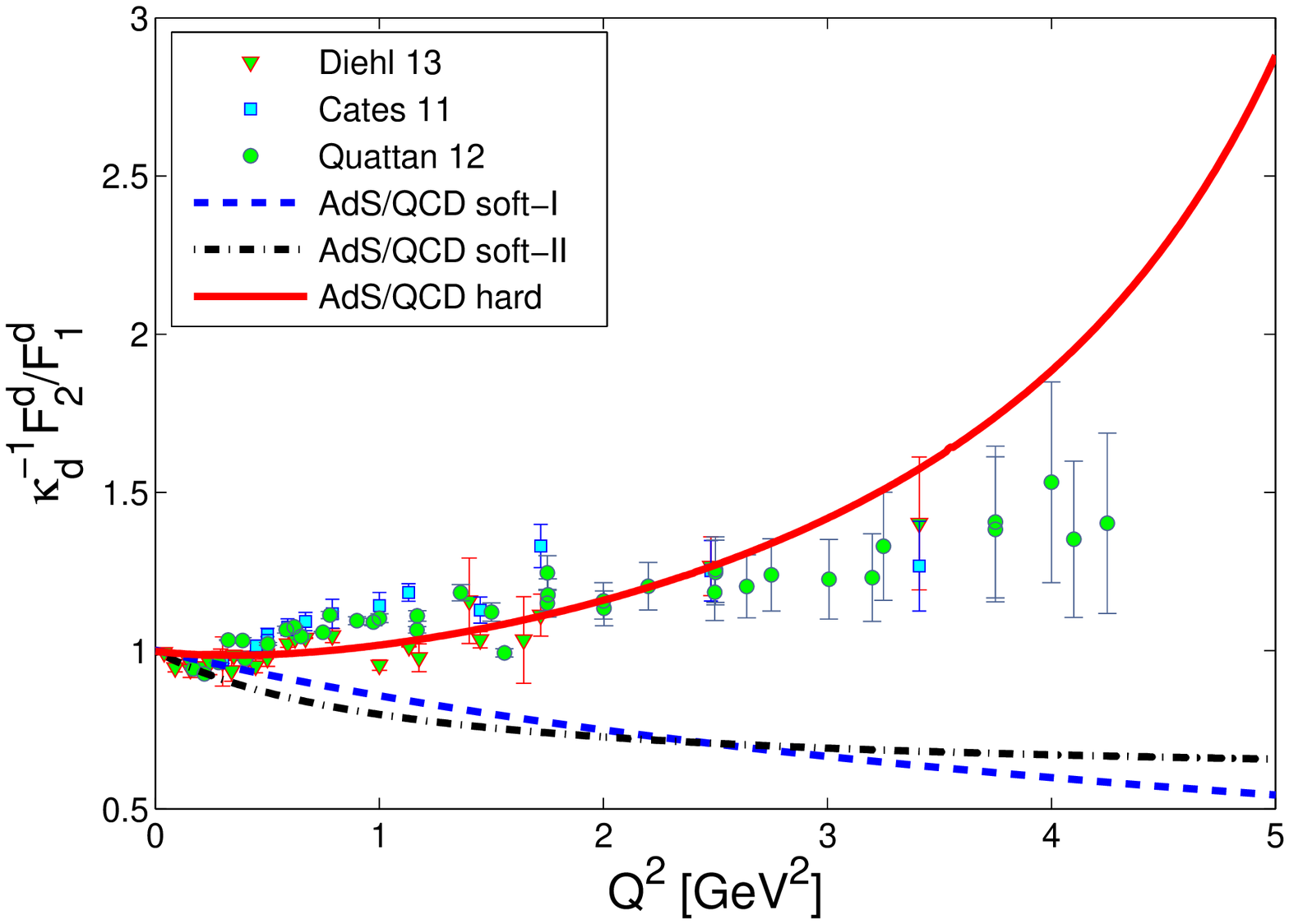}
\caption{\label{flavor_ratio} $F_2^q/{(\kappa_q F_1^q)}$ plotted against $Q^2$ (a) for the $u$ quark and (b) for the $d$ quark.  
}
\end{figure*}
%%%%%%%%%%%%%%%%%%%%%%%%%%%%%%%%%%%%%%%%%%%%%%%%%%%%%%%%%%%%%%%%%%%%%%
\begin{figure*}[htbp]
\small{(a)}\includegraphics[width=7.3cm,height=5.8cm,clip]{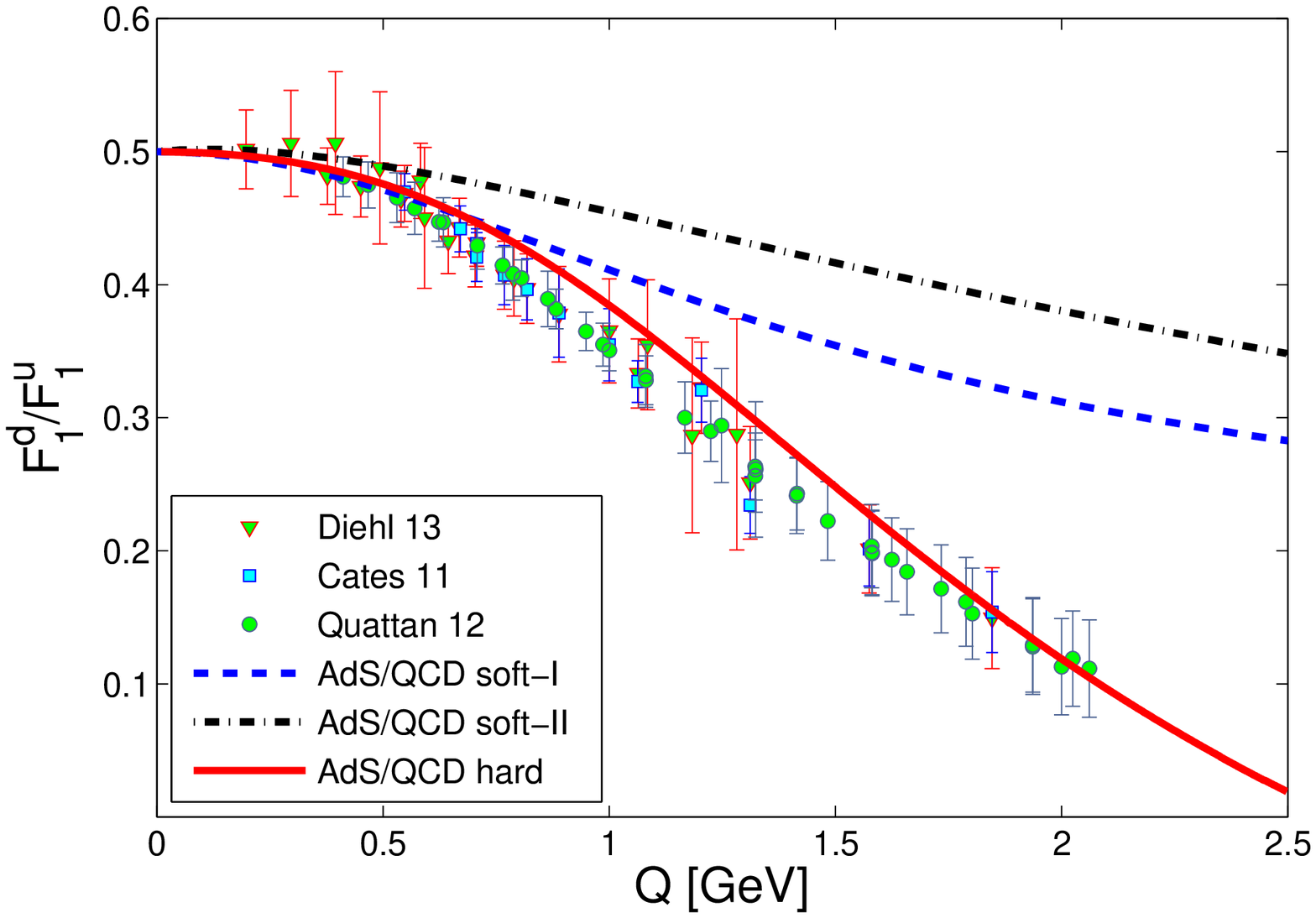}
\small{(b)}\includegraphics[width=7.3cm,height=5.8cm, clip]{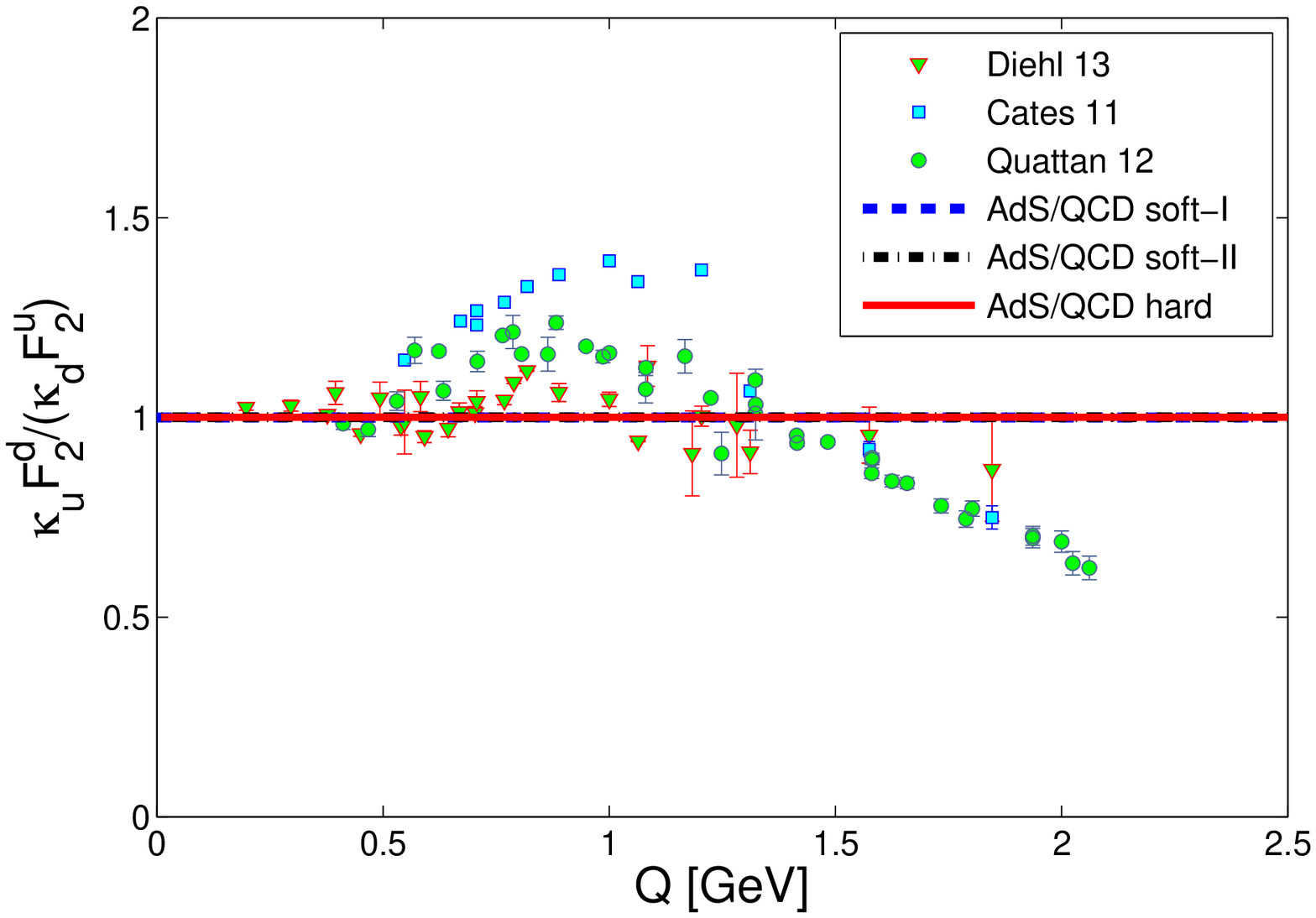}
\caption{\label{flavor_ratio2} The ratios of flavor-dependent form factors (a) $ F_1^d/F_1^u$ and (b) $\kappa_uF_2^d/{\kappa_d F_2^u}$ plotted against $Q=\sqrt{-t}$.
}
\end{figure*}
%%%%%%%%%%%%%%%%%%%%%%
%%%%%%%%%%%%%%%%%%%%%%%%%%%%%%%%%%%%%%%%%%%%%%%%%%%%%%%%%%%%%%%%%%%%%%%%%%
\begin{figure*}[htbp]
\small{(a)}\includegraphics[width=7.3cm,height=5.8cm,clip]{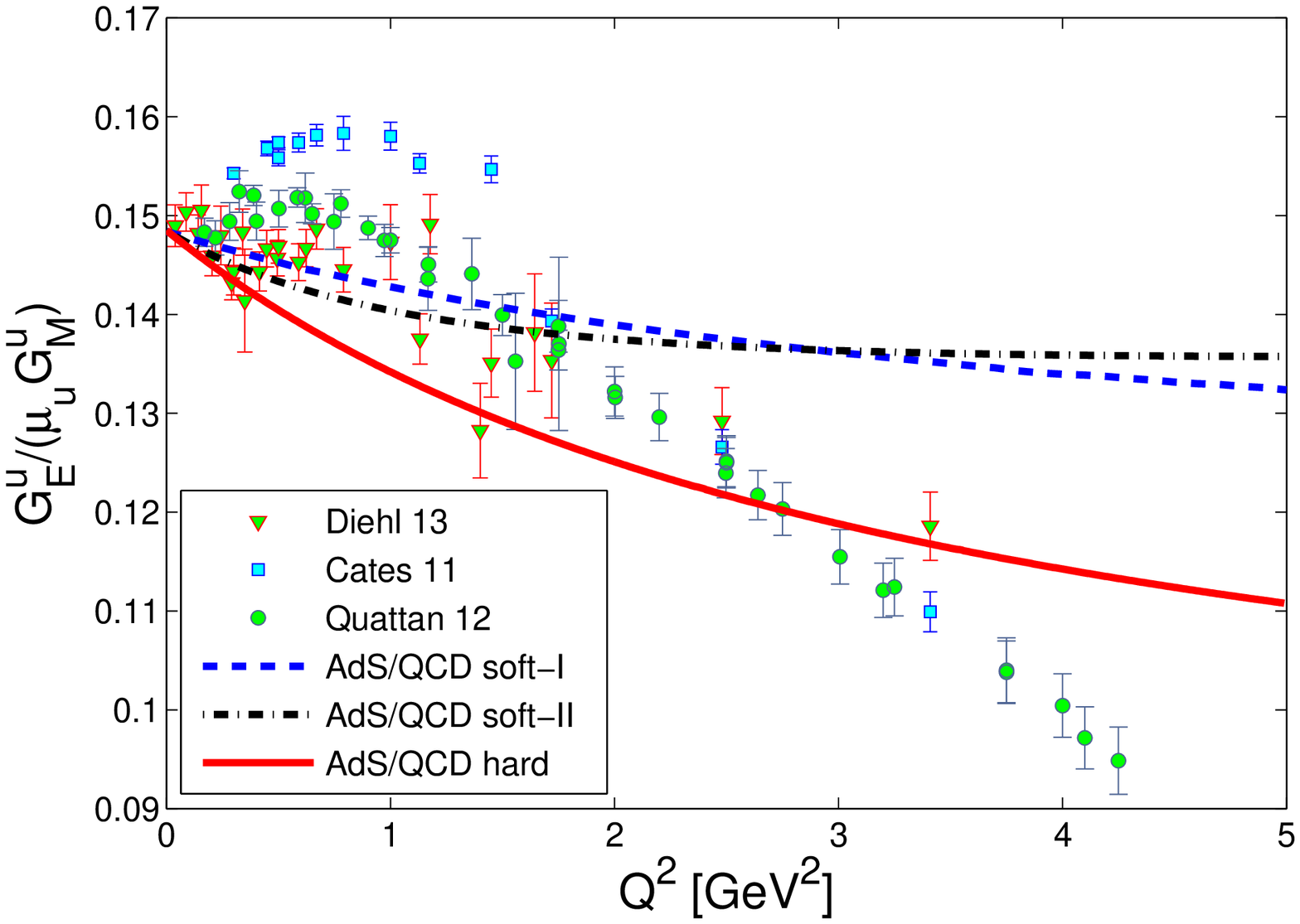}
\small{(b)}\includegraphics[width=7.3cm,height=5.8cm, clip]{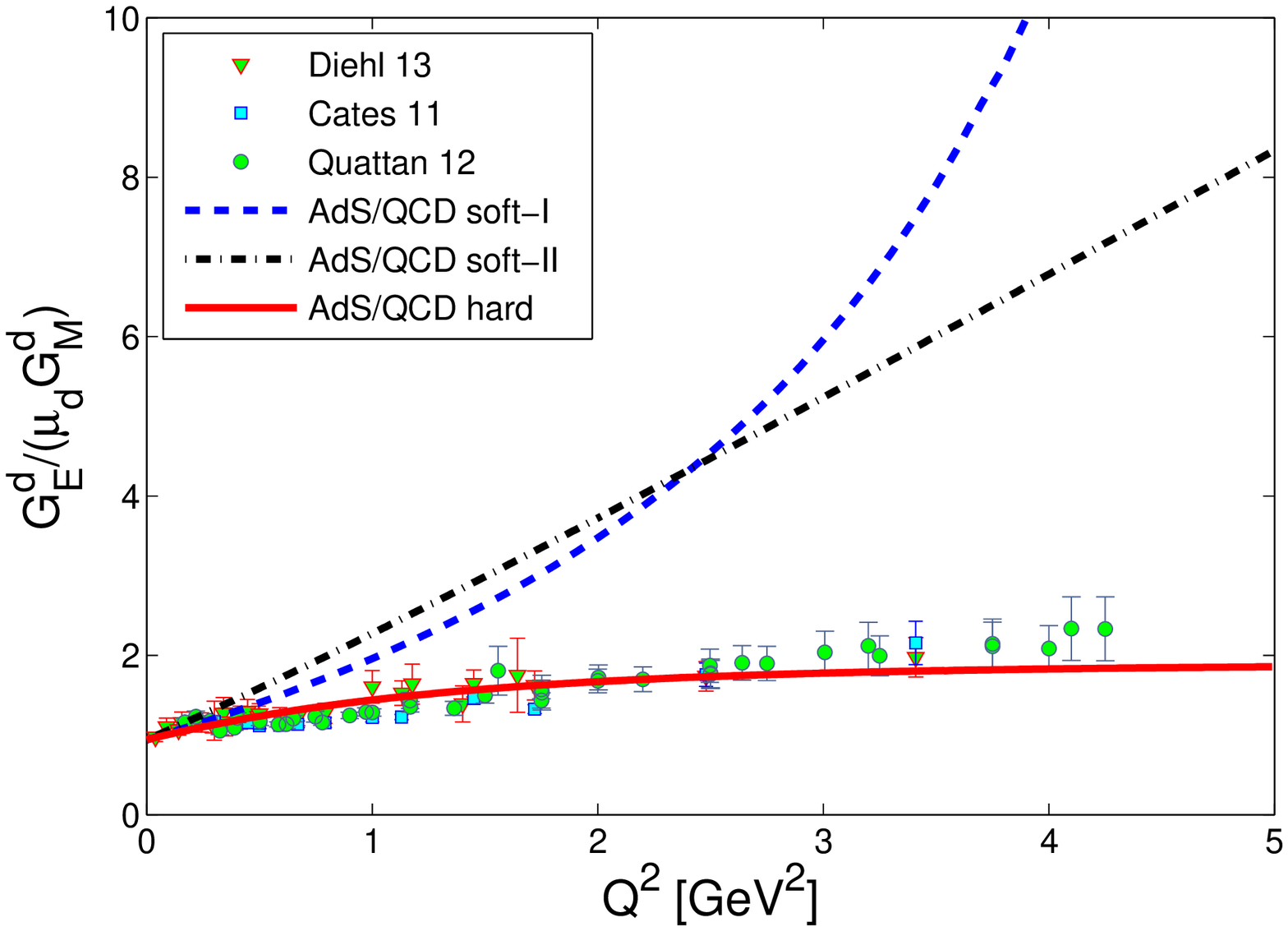}
\caption{\label{Sachs_ratio} $G_E^q/{(\mu_q G_M^q)}$ plotted against $Q^2$ (a) for the $u$ quark and (b) for the $d$ quark.  
}
\end{figure*}
%%%%%%%%%%%%%%%%%%%%%%%%%%%%%%%%%%%%%%%%%%%%%%%%%%%%%%%%%%%%%%%%%%%%%%
The other parameters are fixed from the normalization conditions of the Pauli form factor at $Q^2=0$. They are given by $\eta_p=0.448$ and $\eta_n=-0.478$. The normalizable modes $\psi_L(z)$ and $\psi_R(z)$ are given by \cite{AC}
\be
\psi_L(z)=\frac{\sqrt{2}z^2J_2(m_nz)}{z_0J_2(m_nz_0)}, \quad \quad  \psi_R(z)=\frac{\sqrt{2}z^2J_1(m_nz)}{z_0J_2(m_nz_0)}.
\ee
The bulk-to-boundary propagator for the hard-wall AdS/QCD model is given by \cite{grig}
\be
V(Q,z)=Qz\Big[\frac{K_0(Qz_0)}{I_0(Qz_0)}I_1(Qz)+K_1(Qz)\Big],
\ee
where $J_{\nu}$, $I_{\nu}$, and $K_{\nu}$ are the Bessel and modified Bessel functions.
%%%%%%%%%%%%%%%%%%%%%%%%%%%%%%%%%%%%%%%%%%%%%%%%%%%%%%%%%%%
\subsection{Flavor decompositions of nucleon form factors}
%%%%%%%%%%%%%%%%%%%%%%%%%%%%%%%%%%%%%%%%%%%%%%%%%%%%%%%%%%%
In order to evaluate the flavor form factors, we write the flavor decompositions of the nucleon form factors in a straightforward way using the charge and isospin symmetry \cite{Cates}
\be
F_i^u=2F_i^p+F_i^n ~~{\rm and} ~~F_i^d=F_i^p+2F_i^n,~~(i=1,2)
\ee
with the normalizations $F_1^u(0)=2, F_2^u(0)=\kappa_u$ and $F_1^d(0)=1, F_2^d(0)=\kappa_d$. The anomalous magnetic moments for the $u$ and the $d$ quarks are $\kappa_u=2\kappa_p+\kappa_n=1.673$ and $\kappa_d=\kappa_p+2\kappa_n=-2.033$.
One can also define the Sachs form factors for the quarks in the same way as Dirac and Pauli form factors 
 \be
 G_{E,M}^p&=& e_u G_{E,M}^u+e_d G_{E,M}^d,\nonumber\\
 G_{E,M}^n&=& e_u G_{E,M}^d+e_d G_{E,M}^u,
 \ee
 where $e_q$ denotes the charge of quark $q$; thus, $G_{E,M}^u=2 G_{E,M}^p+G_{E,M}^n$ and  $G_{E,M}^d=G_{E,M}^p+2 G_{E,M}^n$, the $Q^2=0$ values of these form factors $G_E^u(0)=2$, $G_E^d(0)=1$, and the magnetic moments are $G_M^u(0)=\mu_u=(2\mu_p+\mu_n)=3.67\mu_N$, $G_M^d(0)=\mu_d=(2\mu_n+\mu_p)=-1.033\mu_N$. The Sachs form factors are expressed in terms of Dirac and Pauli form factors as
 \be
 G_E^{p/n}(Q^2)&=& F_1^{p/n}(Q^2)-\frac{Q^2}{4M^2}F_2^{p/n}(Q^2),\\
 G_M^{p/n}(Q^2)&=& F_1^{p/n}(Q^2)+F_2^{p/n}(Q^2).
 \ee
 Recently, there have been a lot of studies of flavor form factors; Qattan and Arrington \cite{Qattan} have analyzed the flavor decomposition of the form factors using a similar method as \cite{Cates} but included the two photon exchange processes in the Rosenbluth separation. The experimental data for flavor form factors are used to fit the GPDs for up and down quarks  and also estimated the total angular momentum contribution of each flavor by evaluating Ji's sum rule in \cite{diehl13}. In \cite{miller12}, the nucleon and flavor form factors have been studied in a light-front quark-diquark model.
In \cite{harnandez}, the flavor form factors are also discussed using a model for GPDs. 
The flavor decomposition of the nucleon  Sachs form factors in a relativistic quark model based on Goldstone-bon exchange have been studied in \cite{rohrmoser} and compared with the experimental data.  The flavor form factors have also been studied in the SU(3) chiral quark-soliton model in \cite{silva}.

It was shown in \cite{Cates} that the ratio of Pauli and Dirac form factors for the quark $F_2/F_1$ is almost constant, whereas the $Q^2$ dependence  for the ratio of the proton $F_2^p/F_1^p$ is proportional to $1/Q^2$. For $Q^2>1$ GeV$^2$, the experimental data for $F_2^d$ and $F_1^d$ are roughly proportional to $1/Q^4$ but the dropoff of $F_2^u$ and $F_1^u$ is more gradual \cite{Cates}. In Fig. \ref{proton_fit}, we show the ratio of the Pauli and Dirac form factors for the proton calculated in the framework of the hard-wall AdS/QCD model. The Pauli and Dirac form factors for the $u$ and $d$ quarks are shown in Fig. \ref{FF_flavors}. 
One notices that at higher $Q^2$, the flavor form factors in the hard-wall model deviate from the experimental data, and the deviations are larger for the $u$ quark than for the $d$ quark. For $F_1^d$, the hard-wall AdS/QCD model provides a better result than the soft I, but the overall description of soft II at higher $Q^2$ is better compared to both the soft I and the hard-wall AdS/QCD models. On the other hand, for the other three flavor form factors as shown in Figs. \ref{FF_flavors}(a), \ref{FF_flavors}(c), and \ref{FF_flavors}(d), the soft I produces much better data than the hard-wall and soft II models.
We show the ratio of the Pauli and Dirac form factors for each flavor in Fig. \ref{flavor_ratio}. 
The ratios $F_1^d/F_1^u$ and $\kappa_uF_2^d/(\kappa_d F_2^u)$ are shown in Fig. \ref{flavor_ratio2}. 
In Fig. \ref{Sachs_ratio}, we show the ratios of the Sachs form factors $G_E^q/{(\mu_q G_M^q)}$ for the $u$ and $d$ quarks. The plots show that the hard-wall AdS/QCD model reproduces reasonably good data for the ratios $F_1^d/F_1^u$ and $G_E^d/{(\mu_d G_M^d)}$, whereas the soft-wall AdS/QCD models are unable to reproduce good data for the ratios of flavor form factors which involve the $F_1^d$. It should be mentioned here that at low $Q^2$, the description of $F_2^p/F_1^p$ (see Fig.\ref{proton_fit}) in both soft-wall models is better than that of the hard-wall model, but at higher $Q^2$, it is a little better in the hard-wall model.
%%%%%%%%%%%%%%%%%%%%%%%%%%%%%%%%%%%%%%%%%%%%%%%%%%%%%%
\section{Transverse charge and magnetization densities}\label{density}
%%%%%%%%%%%%%%%%%%%%%%%%%%%%%%%%%%%%%%%%%%%%%%%%%%%%%%%
The transverse charge density for an unpolarized nucleon is given by the Fourier transform of the Dirac form factor \cite{miller07,vande}
\be
\rho_{ch}(b)
&=&\int \frac{d^2q_{\perp}}{(2\pi)^2}F_1(q^2)e^{iq_{\perp}.b_{\perp}}\nonumber\\
&=&\int_0^\infty \frac{dQ}{2\pi}QJ_0(Qb)F_1(Q^2),\label{rho_un}
\ee
where the impact parameter $b=|b_{\perp}|$ represents the position from the transverse center of mass of the nucleon, and $J_0$ is the cylindrical zero order Bessel function. 
The charge density for flavor $\rho_{fch}^q$ can also be written as Eq.(\ref{rho_un}) with $F_1$ replaced
by $F_1^q$ . The magnetization density can be defined by a similar fashion to have the formula
\be
\widetilde{\rho}_{M}(b) &= & \int \frac{d^2q_{\perp}}{(2\pi)^2}F_2(q^2)e^{iq_{\perp}.b_{\perp}}\nonumber\\
&=&\int_0^\infty \frac{dQ}{2\pi}QJ_0(Qb)F_2(Q^2),
\ee
whereas
\be
\rho_m(b)= -b\frac{\partial \widetilde{\rho}_M(b)}{\partial b}
=b\int_0^\infty \frac{dQ}{2\pi}Q^2J_1(Qb)F_2(Q^2).
\ee
%%%%%%%%%%%%%%%%%%%%%%%%%%%%%%%%%%%%%%%%%%%%%%%%%%%%%%%%%%%%%%%%%%%%%%%%%%%%%%%%%%%%%%%%%
%%%%%%%%%%%%%%%%%%%%%%%%%%%%%%%%%%%%%%%%%%%%%%%%%%%%%%%%%%%%%%%%%%%%%%%%%%%%%%%%%%%%%%%%%
$\rho_m(b)$ has been interpreted as an anomalous magnetization density \cite{miller10}. Actual experimental data are unavailable for the charge and magnetization densities, as these quantities are not directly measured in experiments. An approximate estimation of $\rho_{ch}(b)$ and $\rho_m(b)$ for the proton has been done from experimental data of the form factor in Ref. \cite{venkat}.
To get the full information about the transverse charge and magnetization densities inside the nucleon, one needs to evaluate these quantities for different quarks. 
%%%%%%%%%%%%%%%%%%%%%%%%%%%%%%%%%%%%%%%%%%%%%%%%%%%%%%%%%%%%%%%%
\begin{figure*}[htbp]
\begin{minipage}[c]{0.98\textwidth}
\small{(a)}
\includegraphics[width=7.3cm,height=5.8cm,clip]{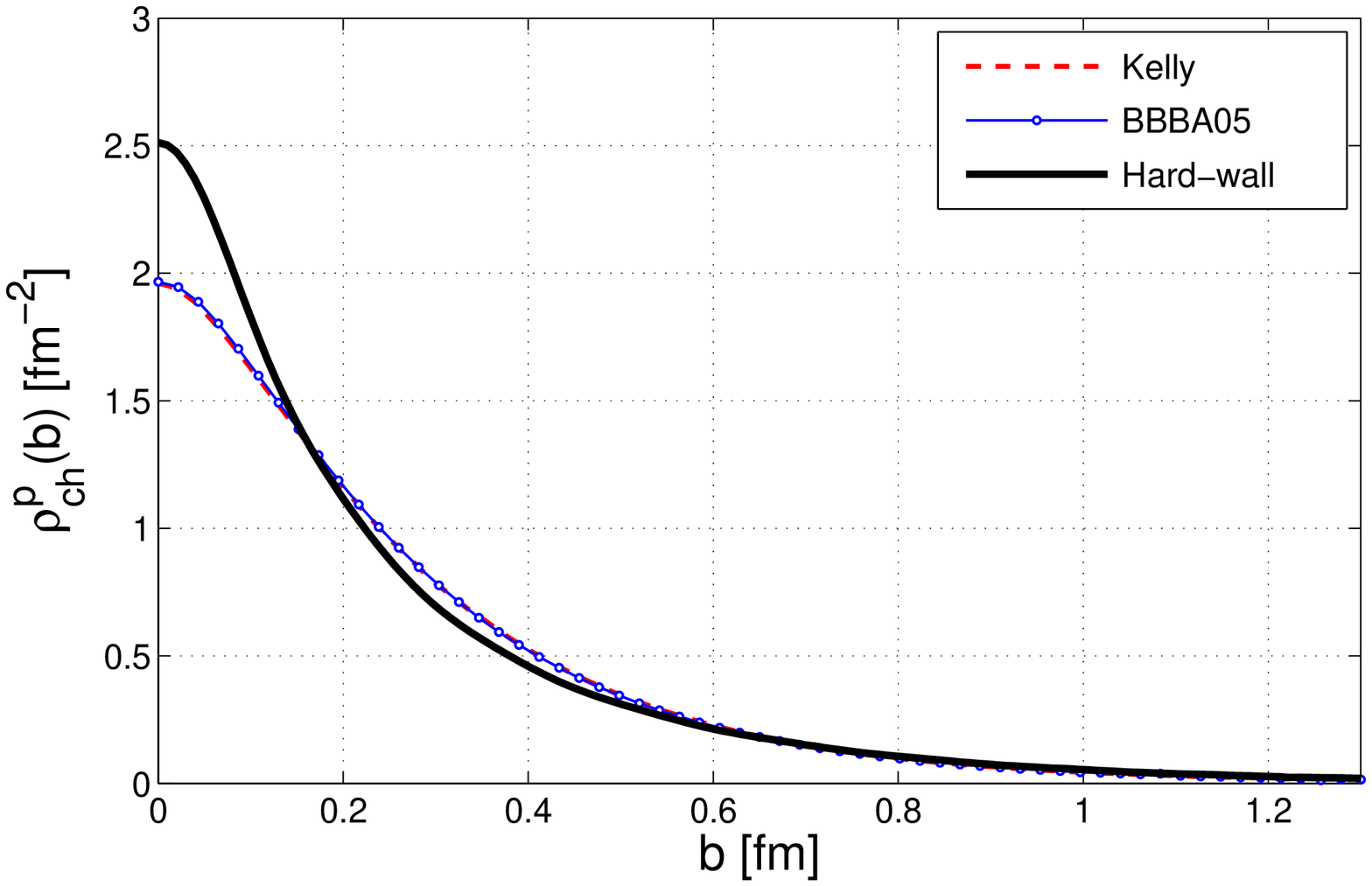}
\hspace{0.1cm}%
\small{(b)}\includegraphics[width=7.3cm,height=5.8cm,clip]{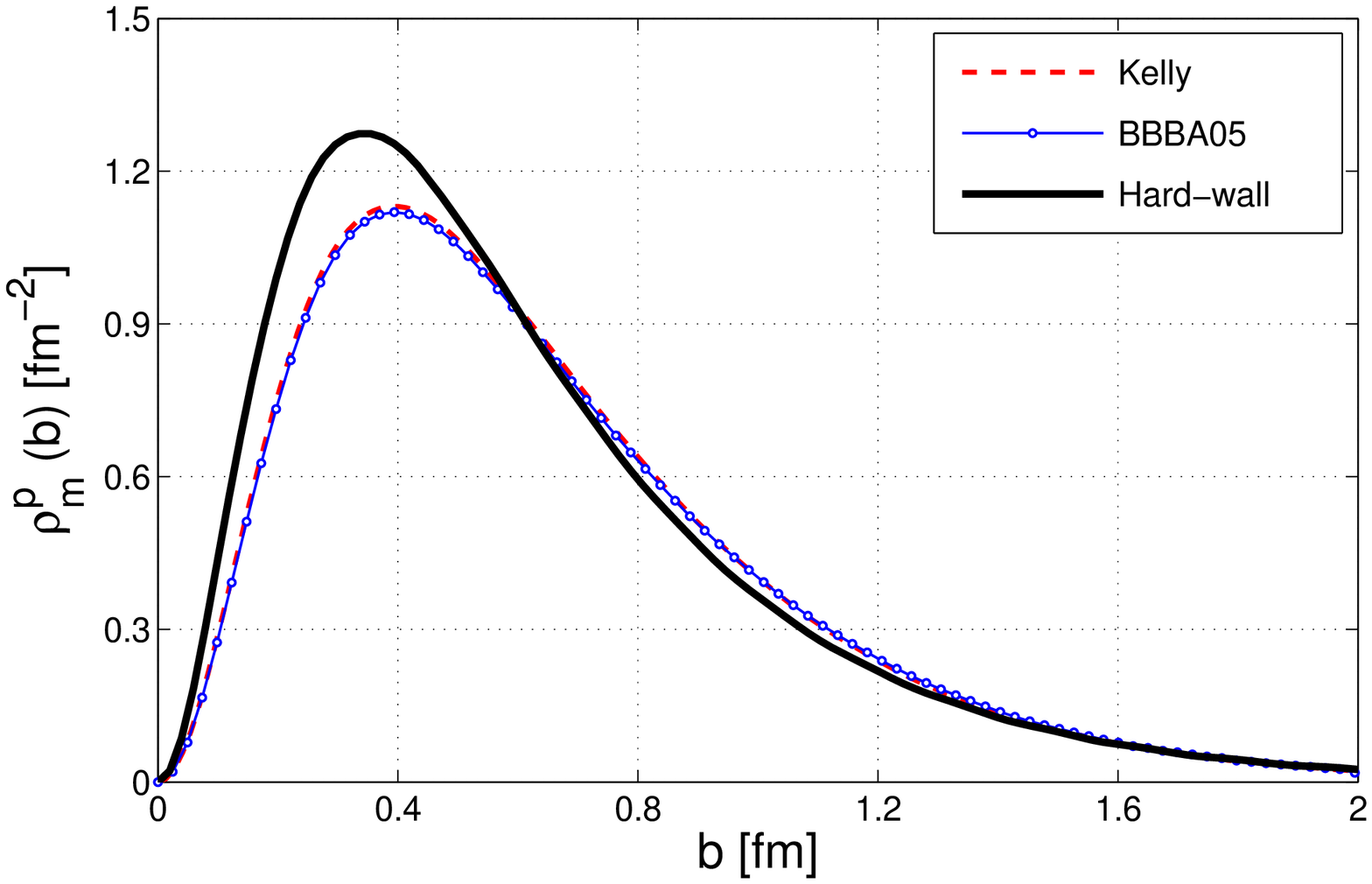}
\end{minipage}
\begin{minipage}[c]{0.98\textwidth}
\small{(c)}\includegraphics[width=7.3cm,height=5.8cm,clip]{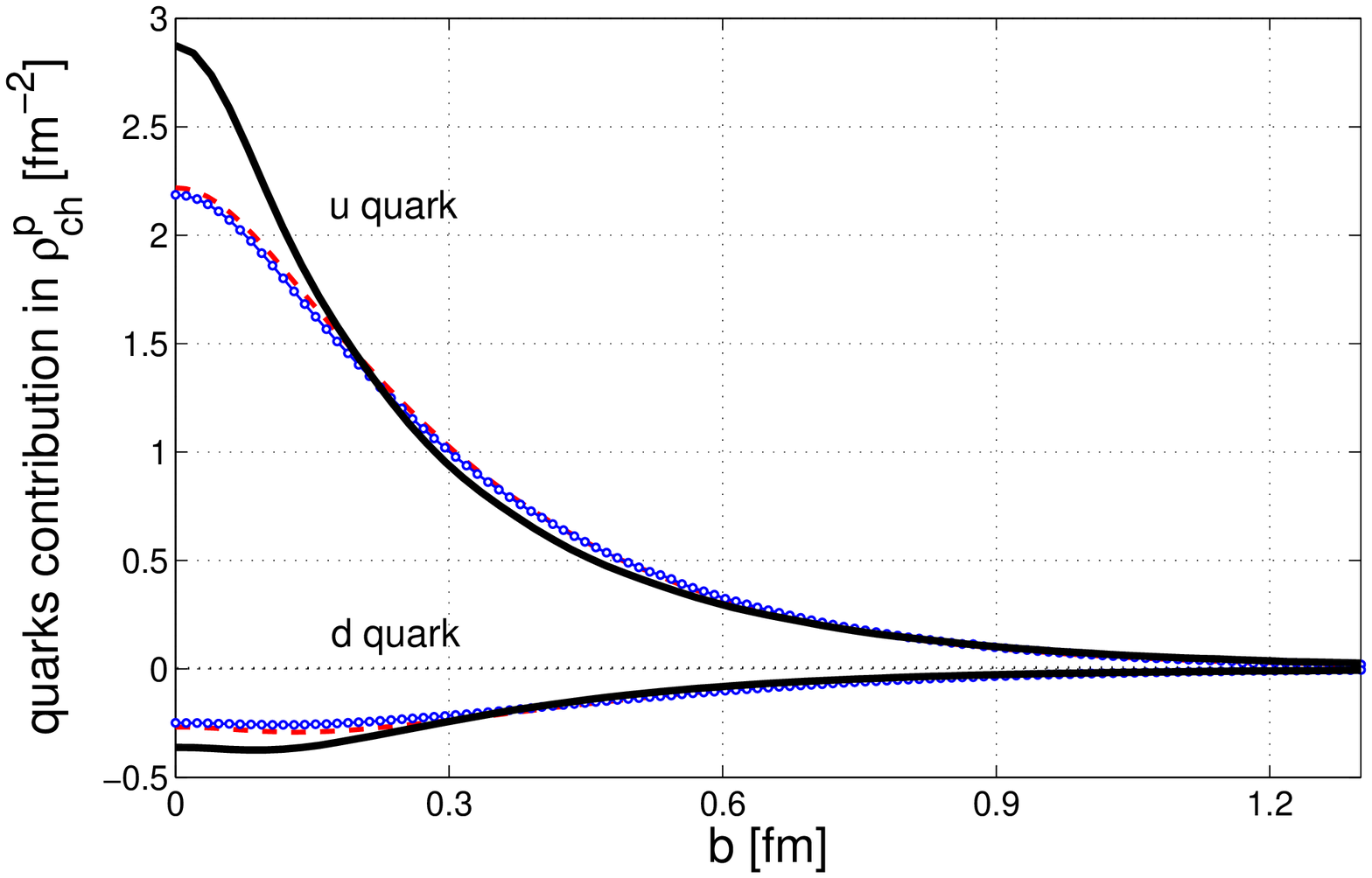}
\hspace{0.1cm}%
\small{(d)}\includegraphics[width=7.3cm,height=5.8cm,clip]{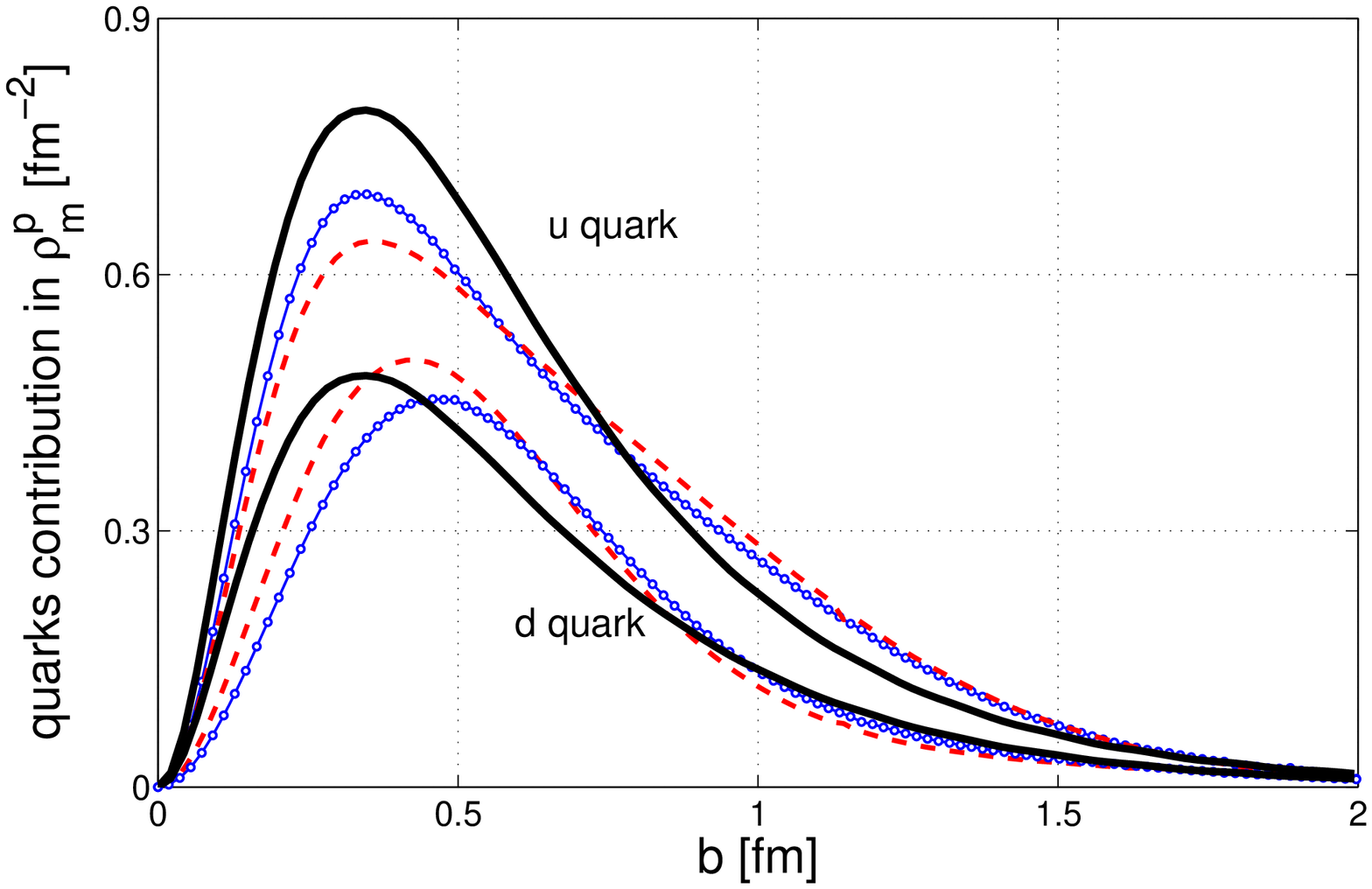}
\end{minipage}
\caption{\label{proton_dencities} Transverse charge and anomalous magnetization densities for the unpolarized proton. (a),(b) represent $\rho_{ch}$ and ${\rho}_m$ for the proton. (c),(d) represent the flavor contributions to the proton densities. The dashed line represents the parametrization of  Kelly \cite{kelly04},  and  the line with circles represents the parametrization of  Bradford $et~al$ \cite{brad}; the solid line is for the hard-wall AdS/QCD model.}
\end{figure*}
%%%%%%%%%%%%%%%%%%%%%%%%%%%%%%%%%%%%%%%%%%%%%%%%%%%%%%%%%%%%%%%%%%%%%%%%%%%%%%%%%%%%%%%%%
%%%%%%%%%%%%%%%%%%%%%%%%%%%%%%%%%%%%%%%%%%%%%%%%%%%%%%%%%%%%%%%%
\begin{figure*}[htbp]
\begin{minipage}[c]{0.98\textwidth}
\small{(a)}
\includegraphics[width=7.3cm,height=5.8cm,clip]{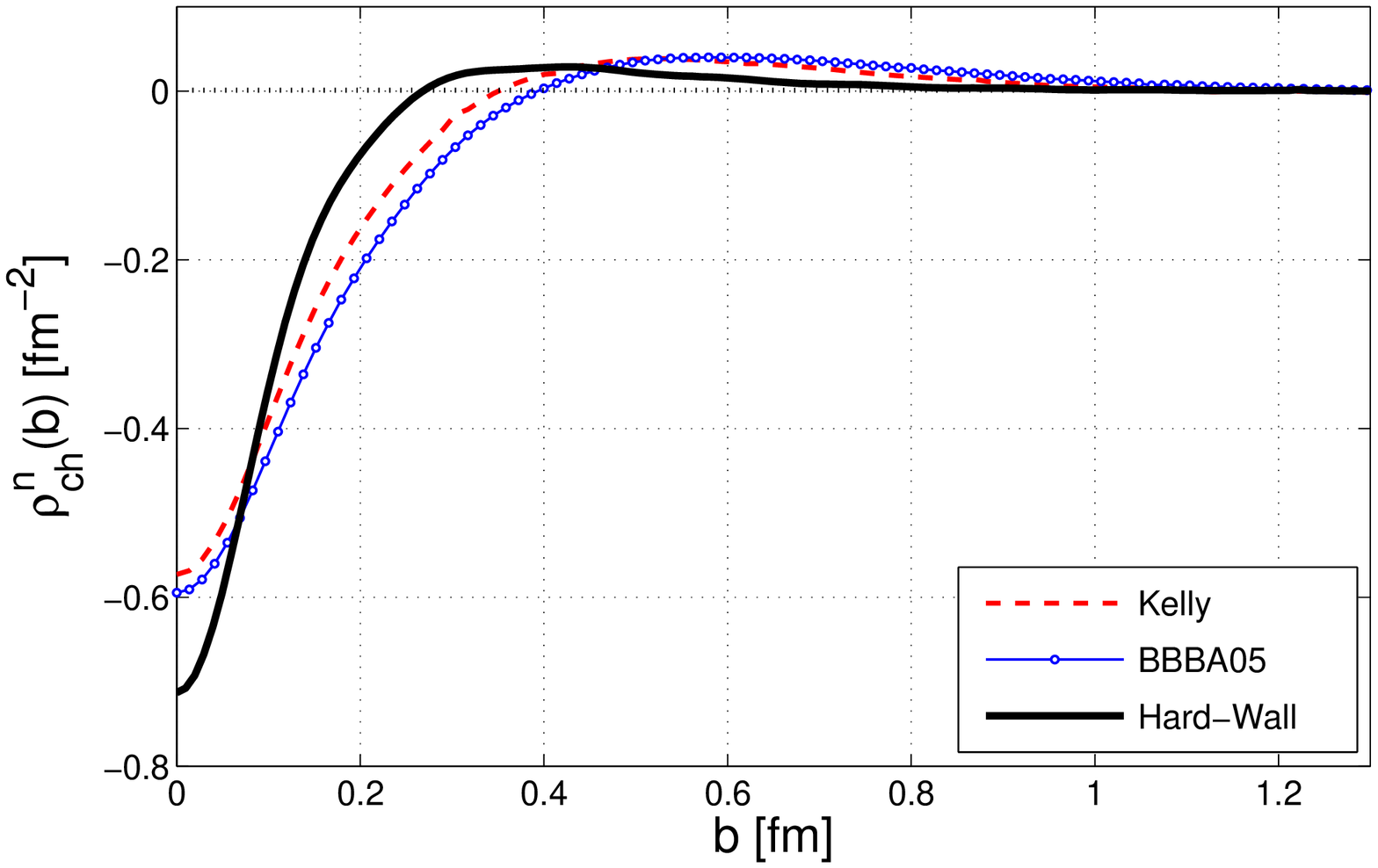}
\hspace{0.1cm}%
\small{(b)}\includegraphics[width=7.3cm,height=5.8cm,clip]{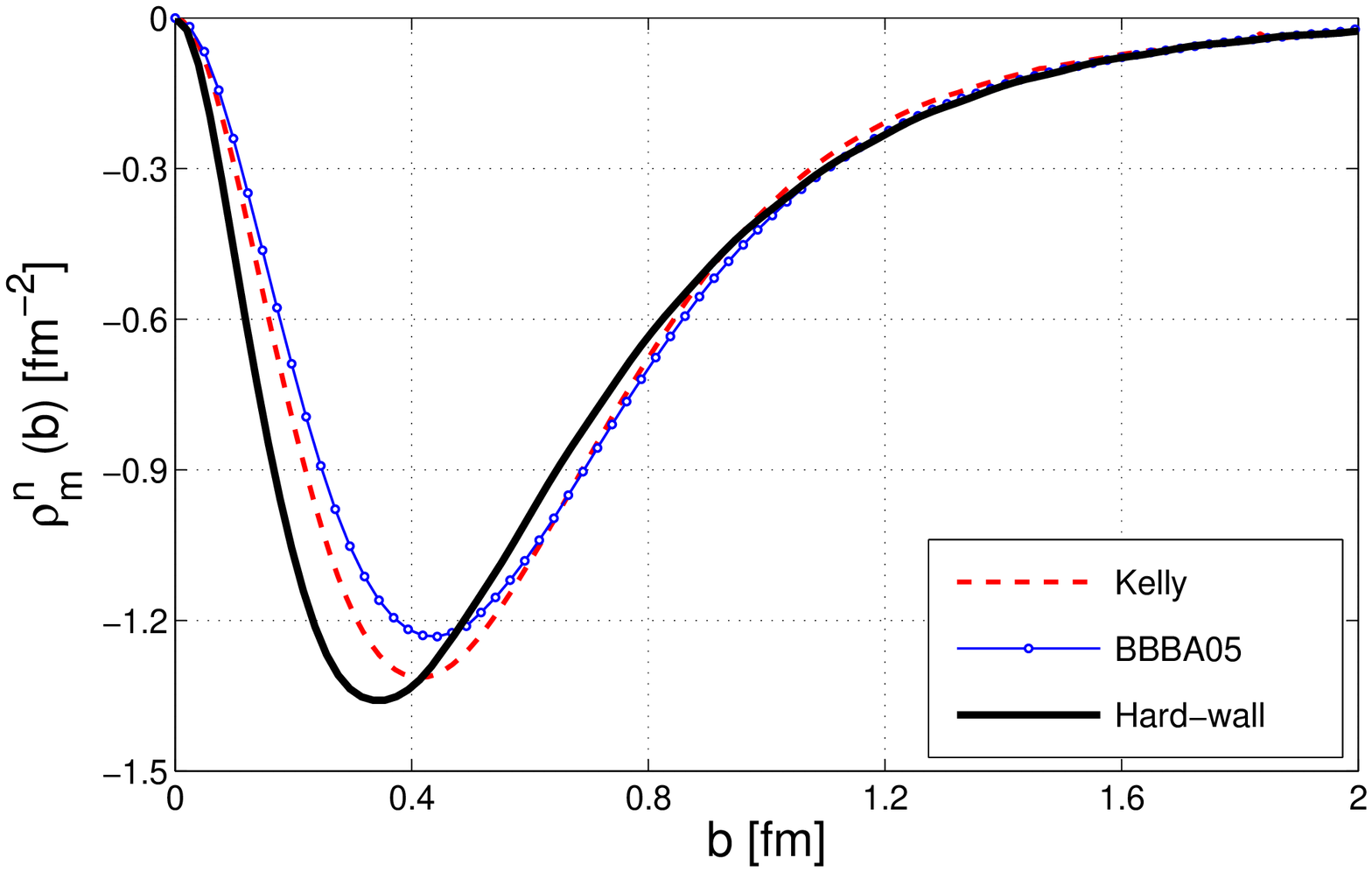}
\end{minipage}
\begin{minipage}[c]{0.98\textwidth}
\small{(c)}\includegraphics[width=7.3cm,height=5.8cm,clip]{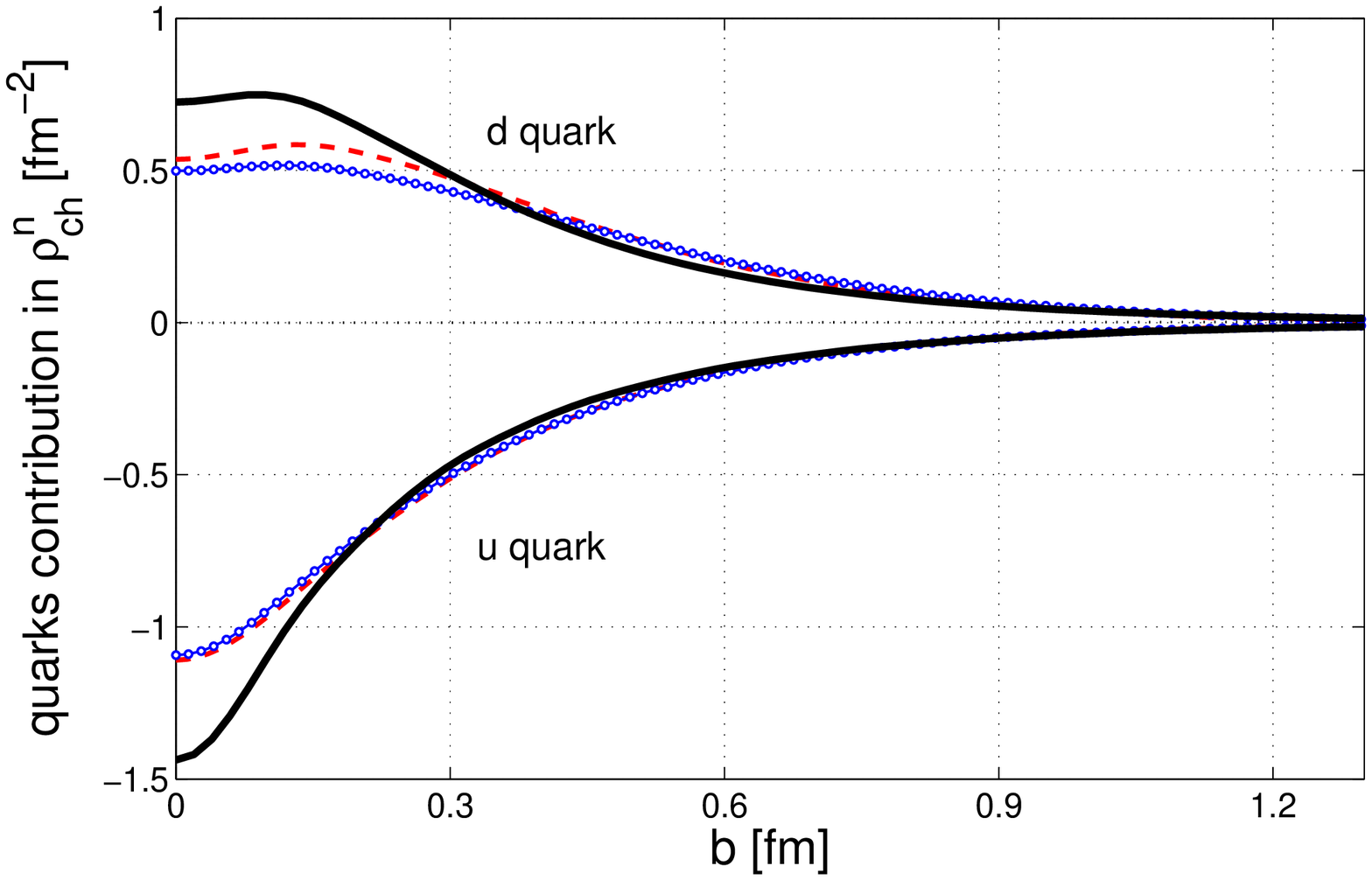}
\hspace{0.1cm}%
\small{(d)}\includegraphics[width=7.3cm,height=5.8cm,clip]{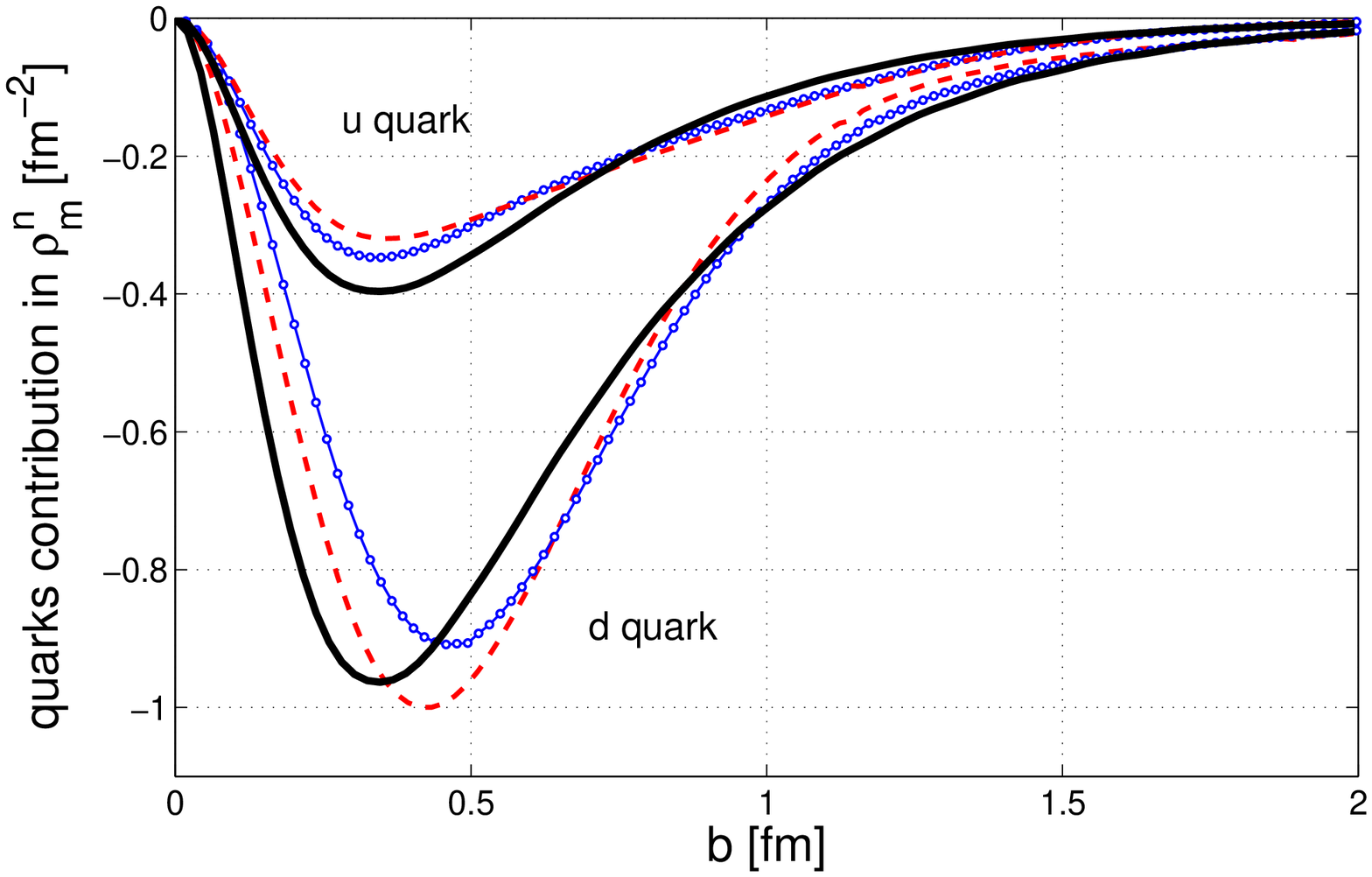}
\end{minipage}
\caption{\label{neutron_densities} Transverse charge and anomalous magnetization densities for the unpolarized neutron. (a),(b) represent $\rho_{ch}$ and ${\rho}_m$ for the neutron. (c),(d) represent the flavor contributions to the neutron densities. The dashed line represents the parametrization of  Kelly \cite{kelly04}, and the line with circles represents the parametrization of  Bradford $et~al$ \cite{brad}; the solid line is for the hard-wall AdS/QCD model.}
\end{figure*}
%%%%%%%%%%%%%%%%%%%%%%%%%%%%%%%
%%%%%%%%%%%%%%%%%%%%%%%%%%%%%%%%%%%%%%%%%%%%%%%%%%%%%%%%%
\subsection{Transverse densities for flavor}
%%%%%%%%%%%%%%%%%%%%%%%%%%%%%%%%%%%%%%%%%%%%%%%%%%%%%%%%%
The decompositions of the transverse charge and magnetization densities for the nucleon can be defined in a similar way as the electromagnetic form factors \cite{Cates}. In terms of two flavors, one can write the charge density decompositions as \cite{CM3}  
\be
 \rho_{ch}^p&=& e_u \rho_{fch}^u+e_d \rho_{fch}^d,\nonumber\\\label{ch_mag1}
 \rho_{ch}^n&=& e_u \rho_{fch}^d+e_d \rho_{fch}^u,
 \ee
where $e_u$ and $e_d$ are the charges of the $u$ and $d$ quarks, respectively. It has been shown in Ref. \cite{miller07} that under the charge and isospin symmetry, the $u$, $d$ quark densities in the proton are the same as the $d$, $u$ densities in the neutron. It is straightforward to write down the transverse charge density for the $u$ and $d$ quark as \cite{CM3, miller07}  
\be
 \rho_{ch}^u(b)&=&  \rho_{ch}^p+ \frac{\rho_{ch}^n}{2}=\frac{ \rho_{fch}^u}{2},\nonumber\\\label{ch_mag2}
 \rho_{ch}^d(b)&=&  \rho_{ch}^p+2 \rho_{ch}^n= \rho_{fch}^d,
 \ee
where $\rho^q_{fch}$ is the charge density for each flavor and $\rho_{ch}^q(b)$ is the charge density of each quark. 
One can also decompose the anomalous magnetization density in a similar fashion as the charge density in Eqs.(\ref{ch_mag1}) and (\ref{ch_mag2}).
%%%%%%%%%%%%%%%%%%%%%%
 \begin{figure*}[htbp]
\begin{minipage}[c]{0.98\textwidth}
\small{(a)}
\includegraphics[width=7.3cm,height=5.8cm,clip]{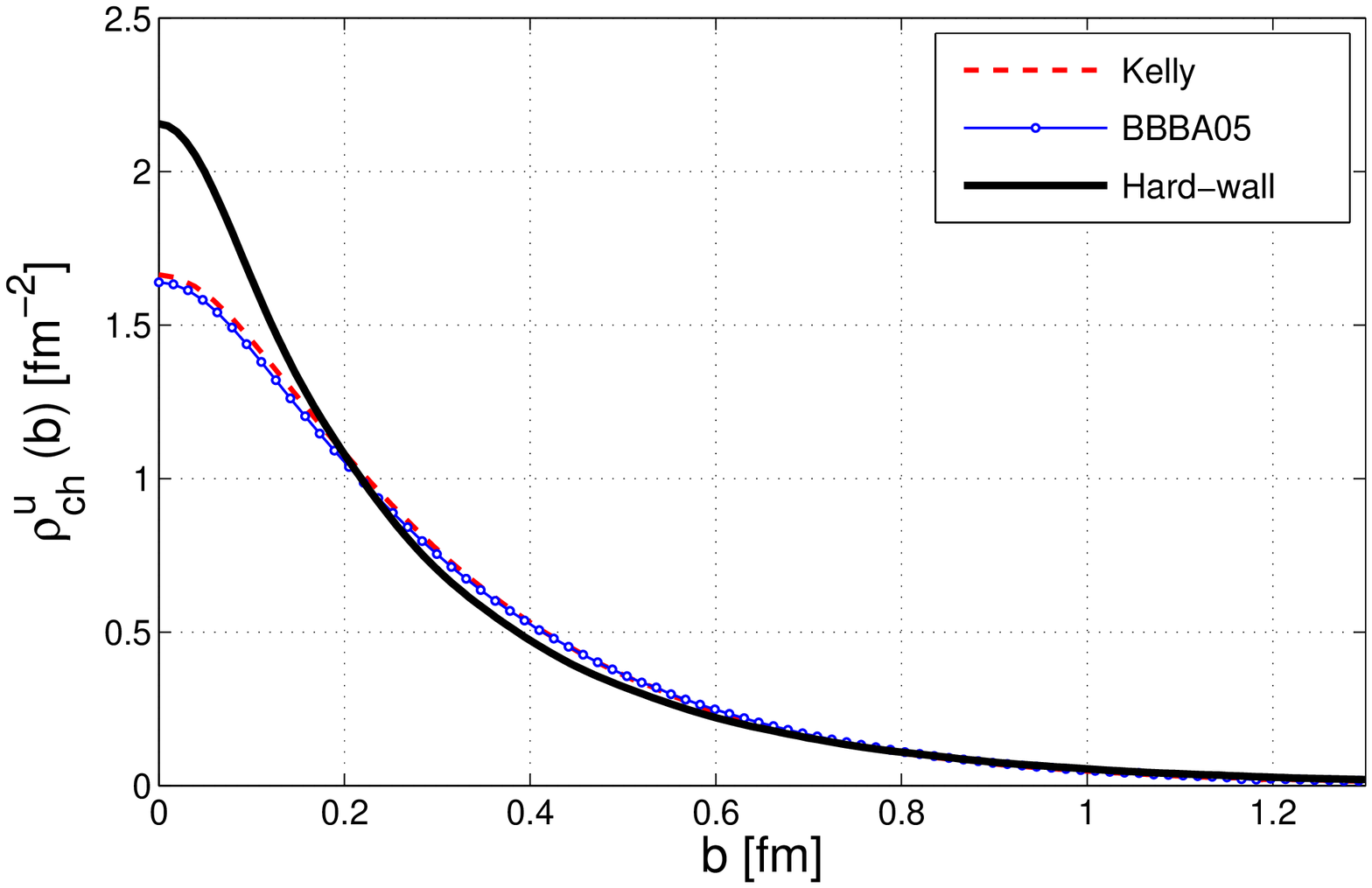}
\hspace{0.1cm}%
\small{(b)}\includegraphics[width=7.3cm,height=5.8cm,clip]{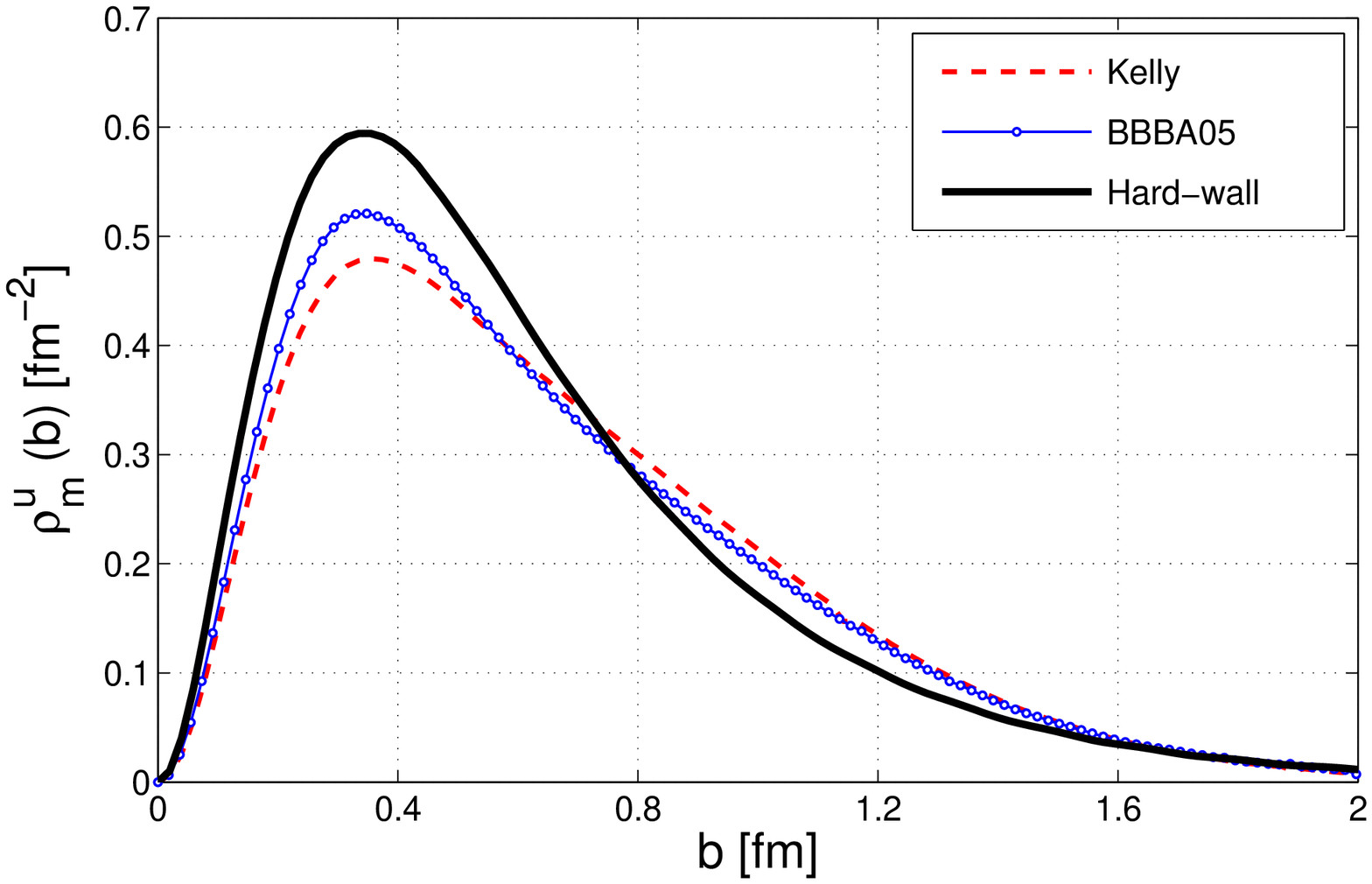}
\end{minipage}
\begin{minipage}[c]{0.98\textwidth}
\small{(c)}\includegraphics[width=7.3cm,height=5.8cm,clip]{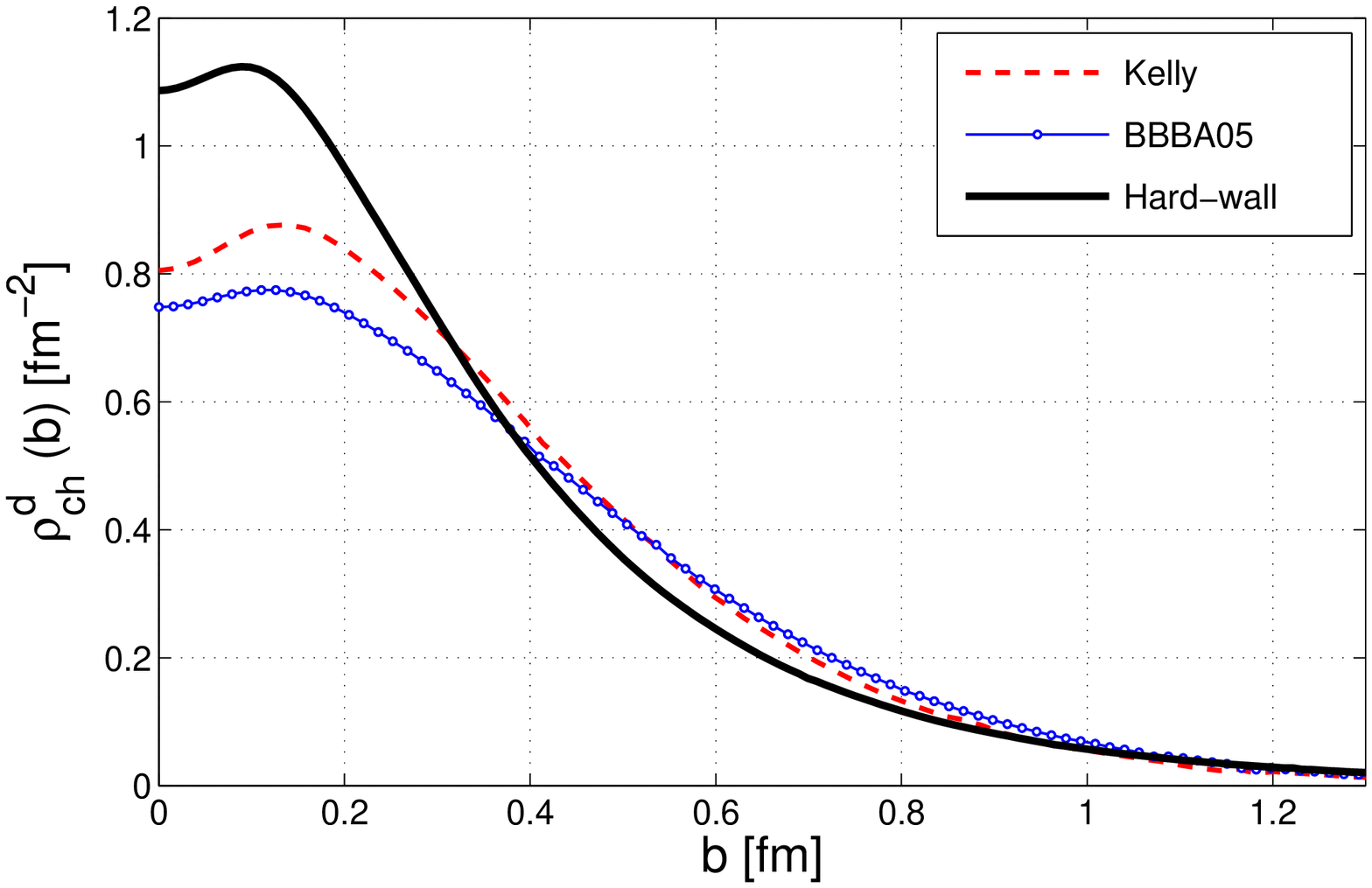}
\hspace{0.1cm}%
\small{(d)}\includegraphics[width=7.3cm,height=5.8cm,clip]{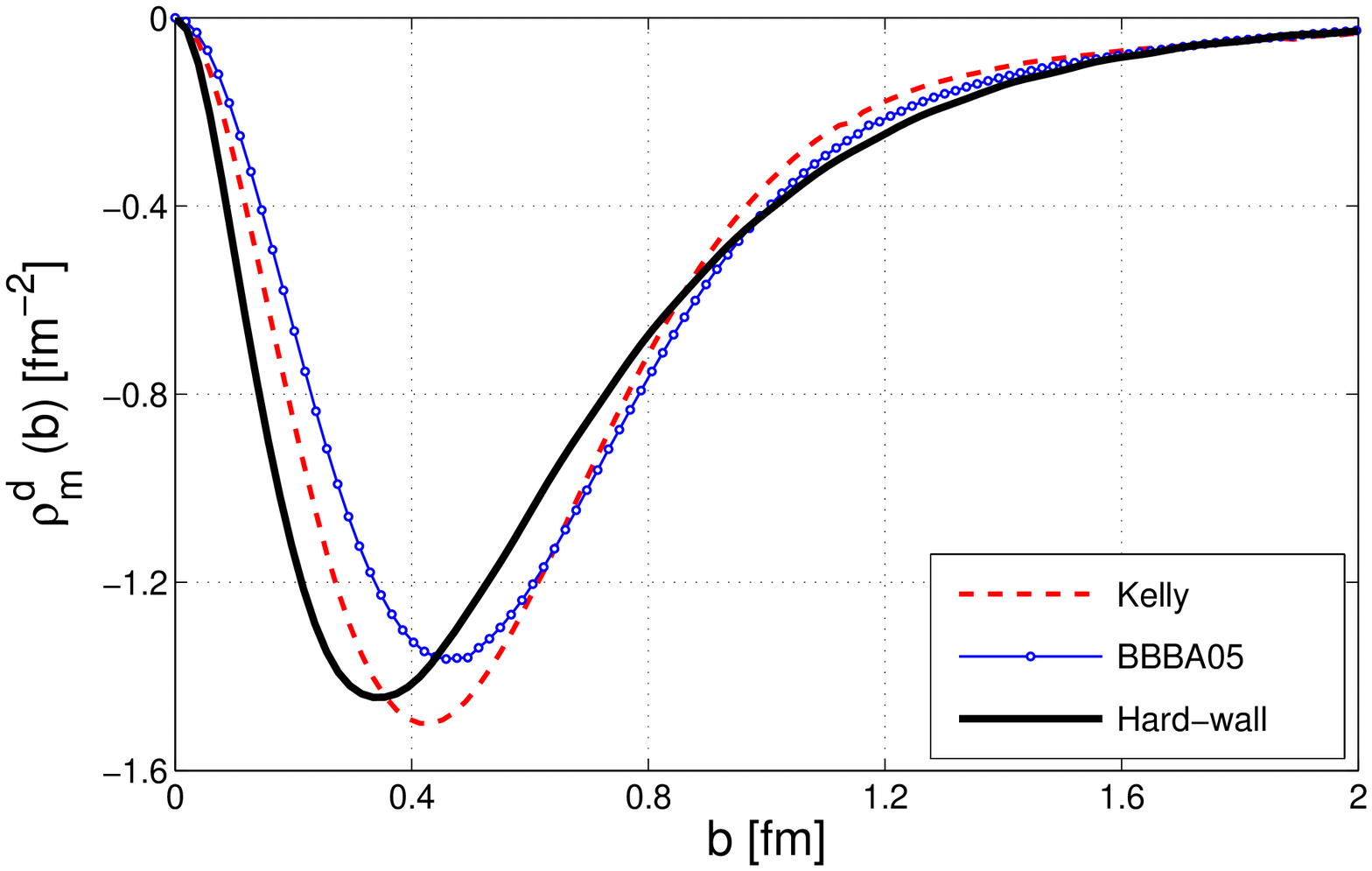}
\end{minipage}
\caption{\label{flavors} Quark transverse charge and anomalous magnetization densities for the unpolarized nucleon. (a),(b) represent $\rho_{ch}$ and ${\rho}_m$ for the $u$ quark. (c),(d) represent the similar densities for the $d$ quark.  Dashed line represents the parametrization of  Kelly \cite{kelly04},  and  the line with circles represents the parametrization of  Bradford $et~al$ \cite{brad}.}
\end{figure*}
%%%%%%%%%%%%%%%
\begin{figure*}[htbp]
\begin{minipage}[c]{0.98\textwidth}
\small{(a)}
\includegraphics[width=7.3cm,height=5.8cm,clip]{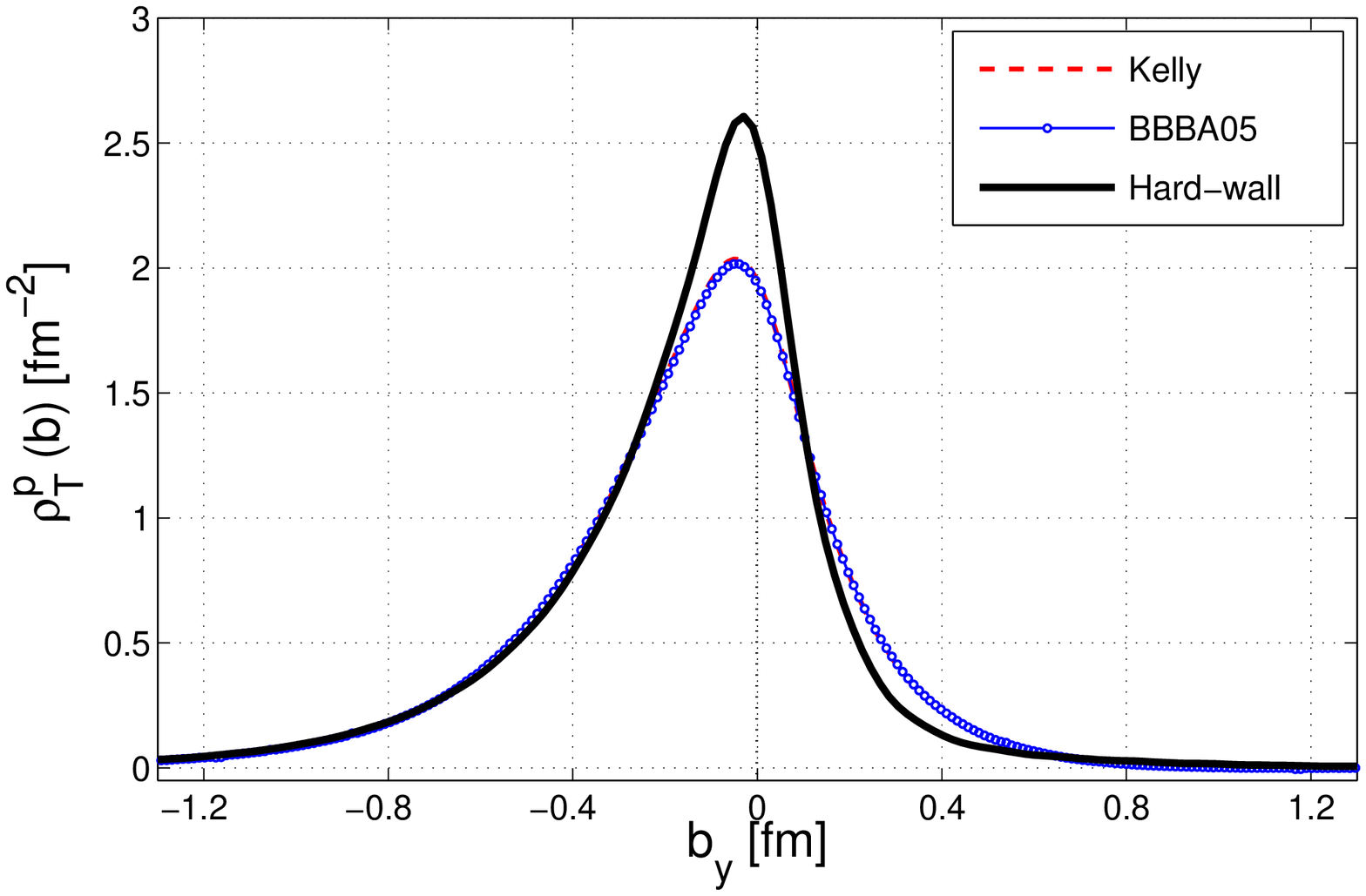}
\hspace{0.1cm}%
\small{(b)}\includegraphics[width=7.3cm,height=5.8cm,clip]{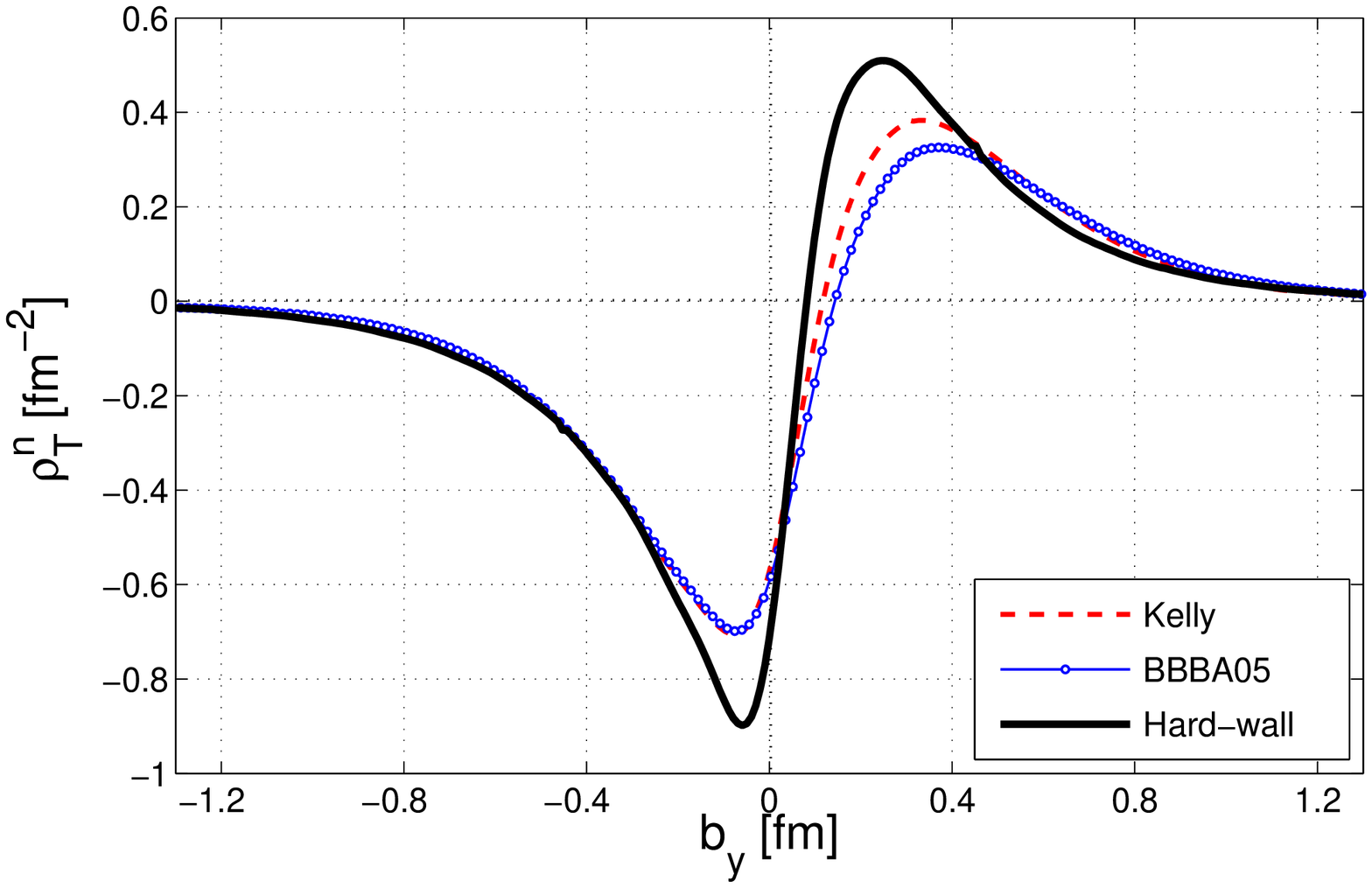}
\end{minipage}
\begin{minipage}[c]{0.98\textwidth}
\small{(c)}
\includegraphics[width=7.3cm,height=5.8cm,clip]{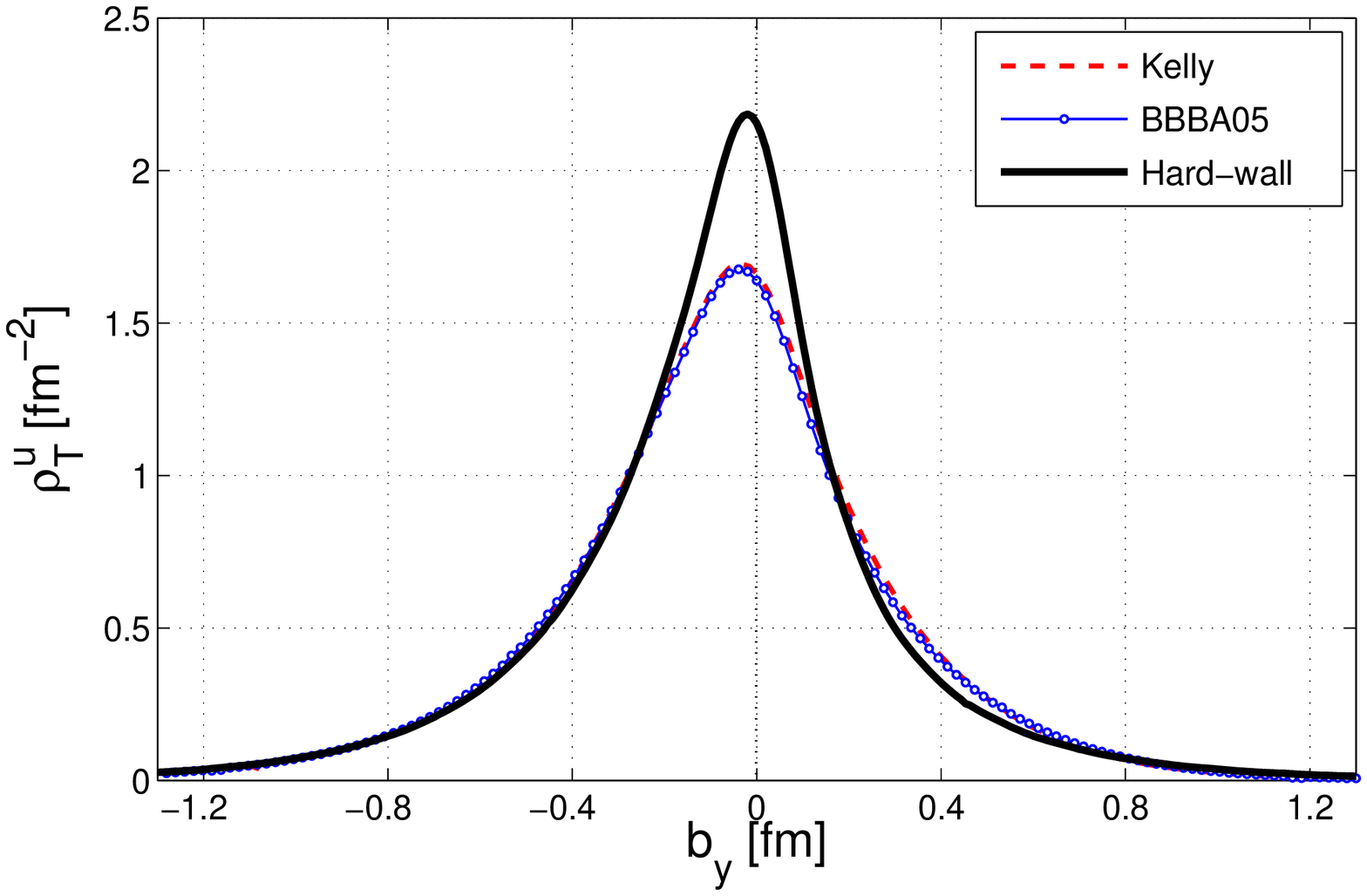}
\hspace{0.1cm}%
\small{(d)}\includegraphics[width=7.3cm,height=5.8cm,clip]{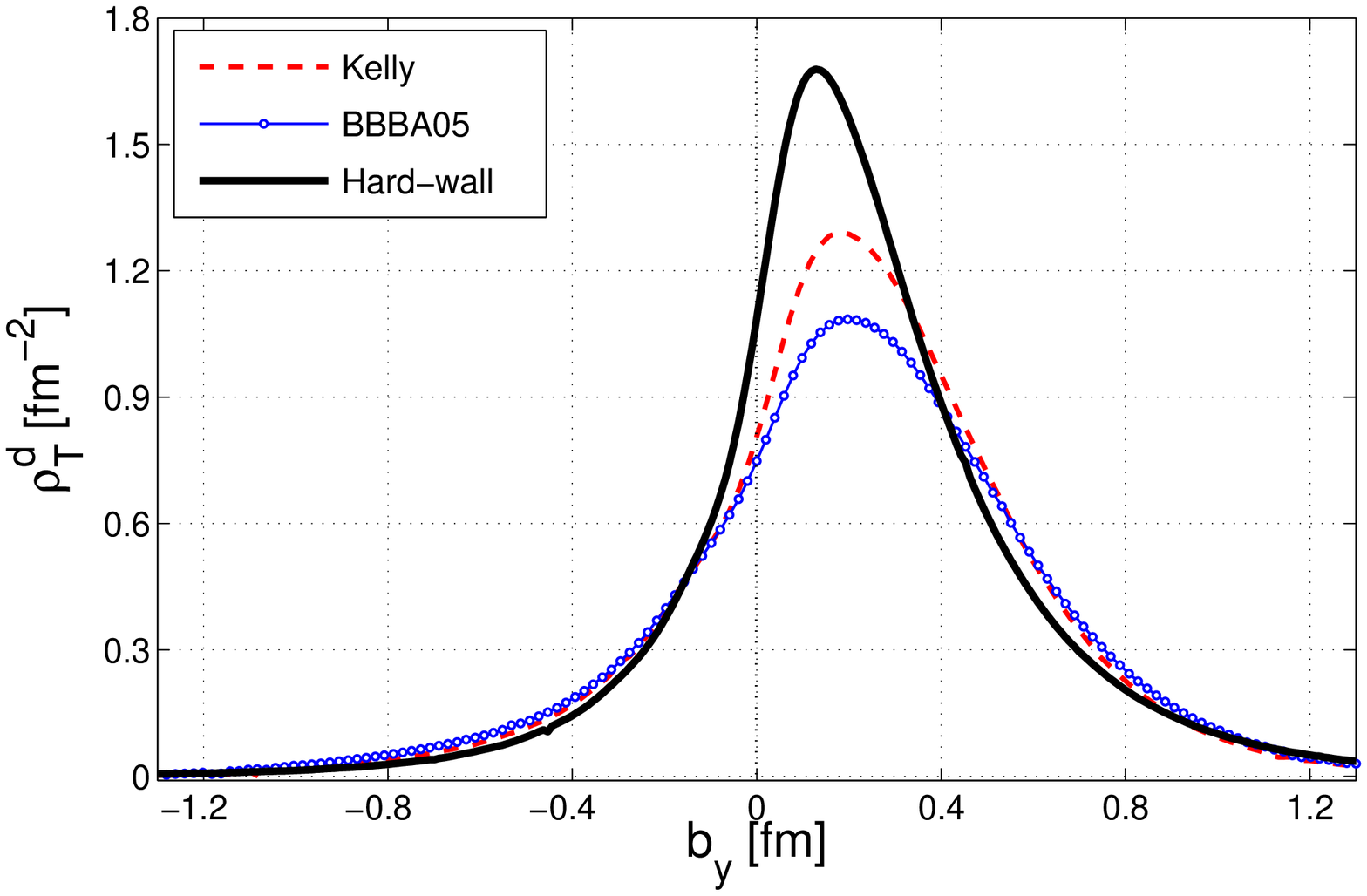}
\end{minipage}
\caption{\label{flavor_T} The charge  densities for the transversely polarized (a) proton, (b) neutron, (c) up, and (d) down quark charge  densities for the transversely polarized nucleon. 
Dashed line represents the parametrization of  Kelly \cite{kelly04},  and  line with circles represents the parametrization of  Bradford $et~al$ \cite{brad}.  }
\end{figure*}
%%%%%%%%%%%%%%%%%
\begin{figure*}[htbp]
\begin{minipage}[c]{0.98\textwidth}
\small{(a)}
\includegraphics[width=7.3cm,height=5.8cm,clip]{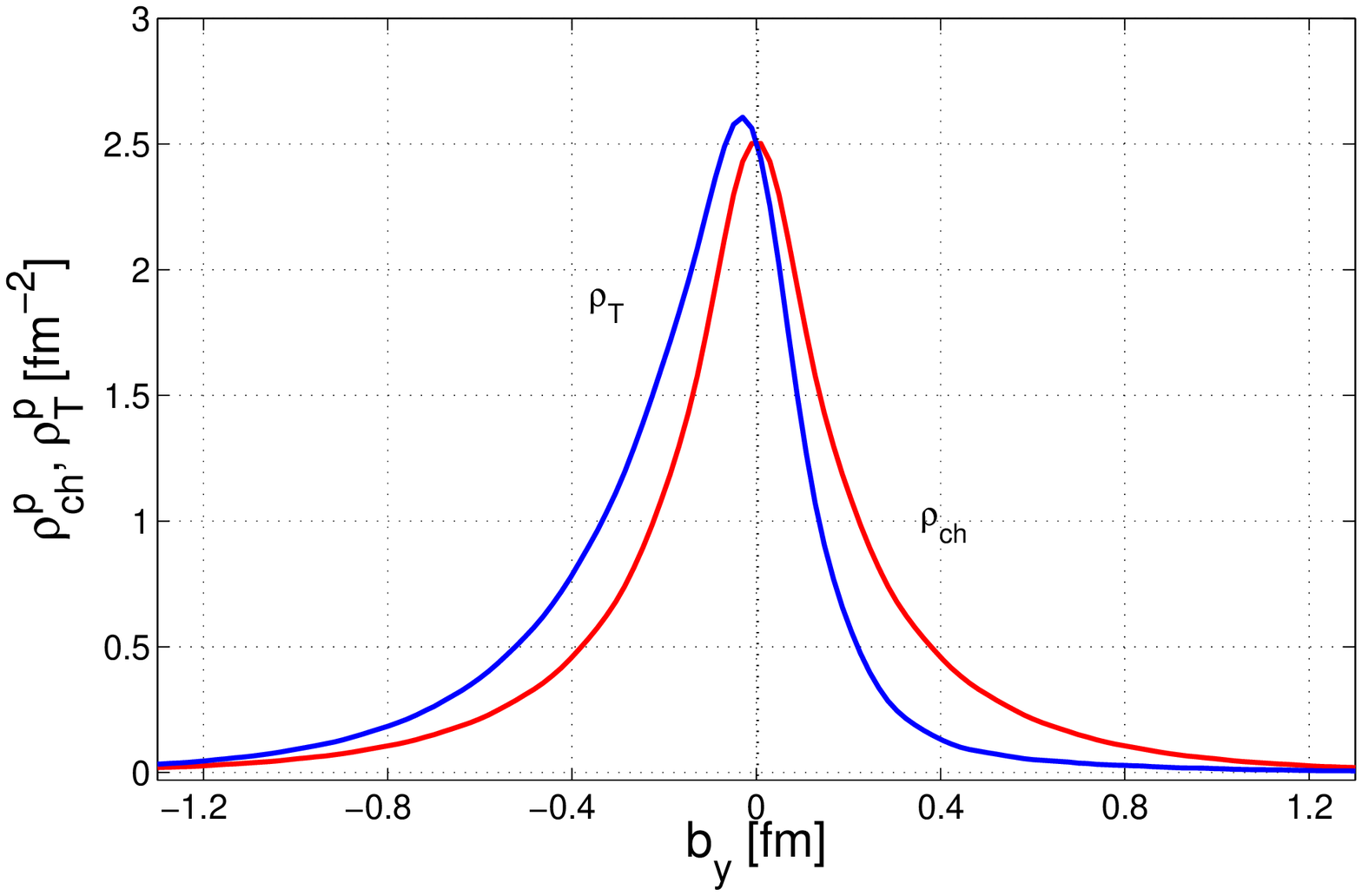}
\hspace{0.1cm}%
\small{(b)}\includegraphics[width=7.3cm,height=5.8cm,clip]{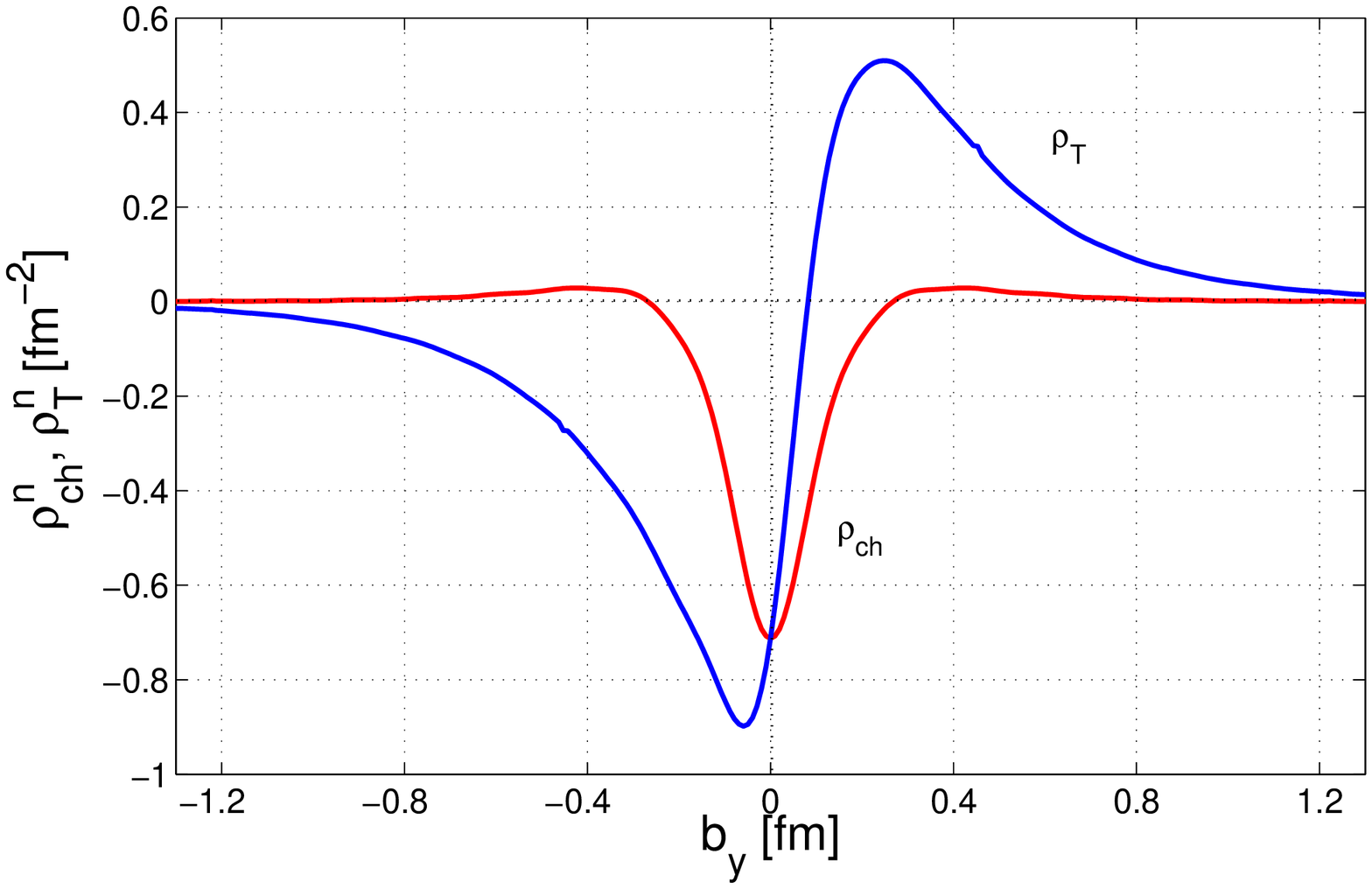}
\end{minipage}
\begin{minipage}[c]{0.98\textwidth}
\small{(c)}
\includegraphics[width=7.3cm,height=5.8cm,clip]{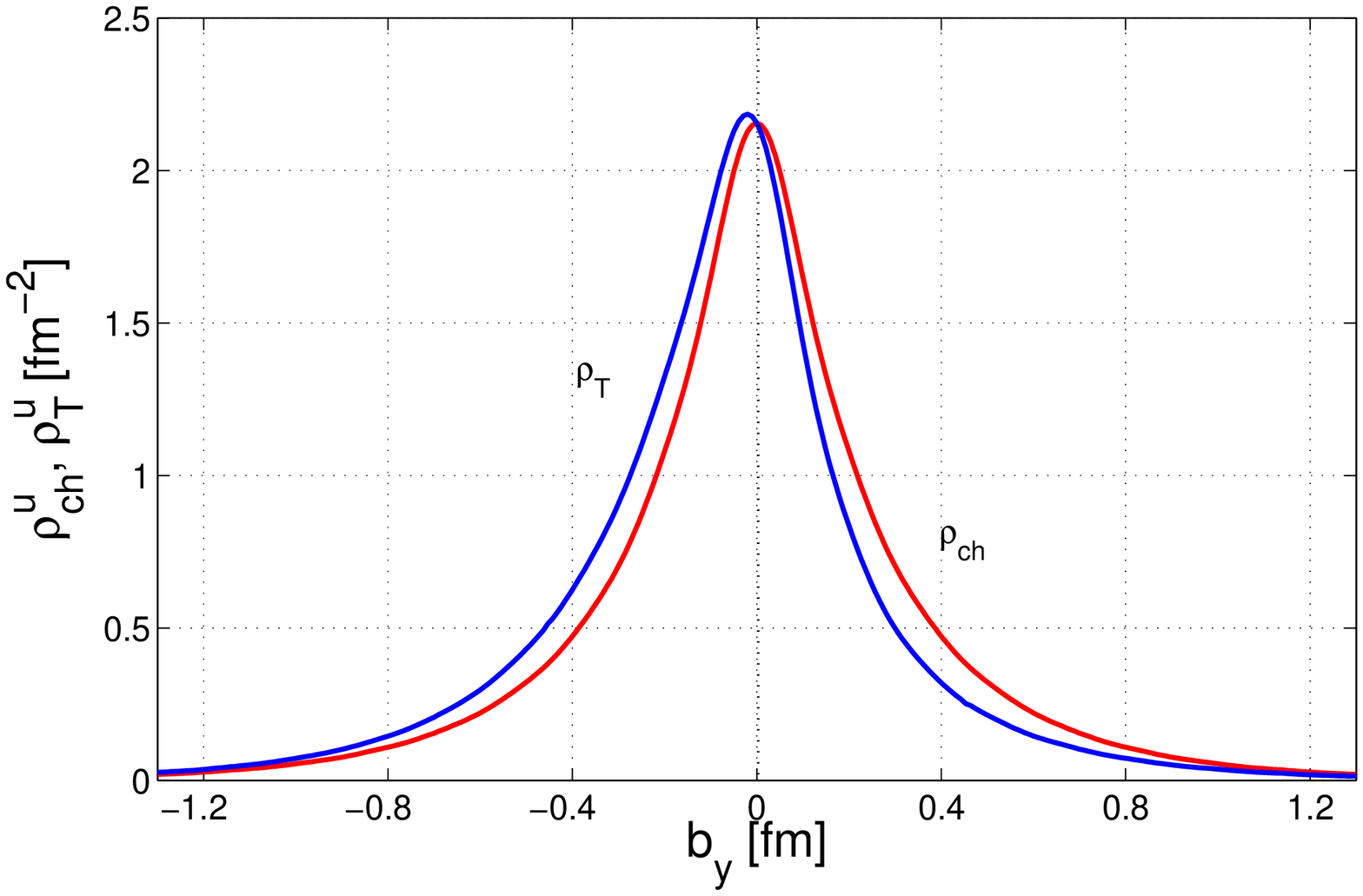}
\hspace{0.1cm}%
\small{(d)}\includegraphics[width=7.3cm,height=5.8cm,clip]{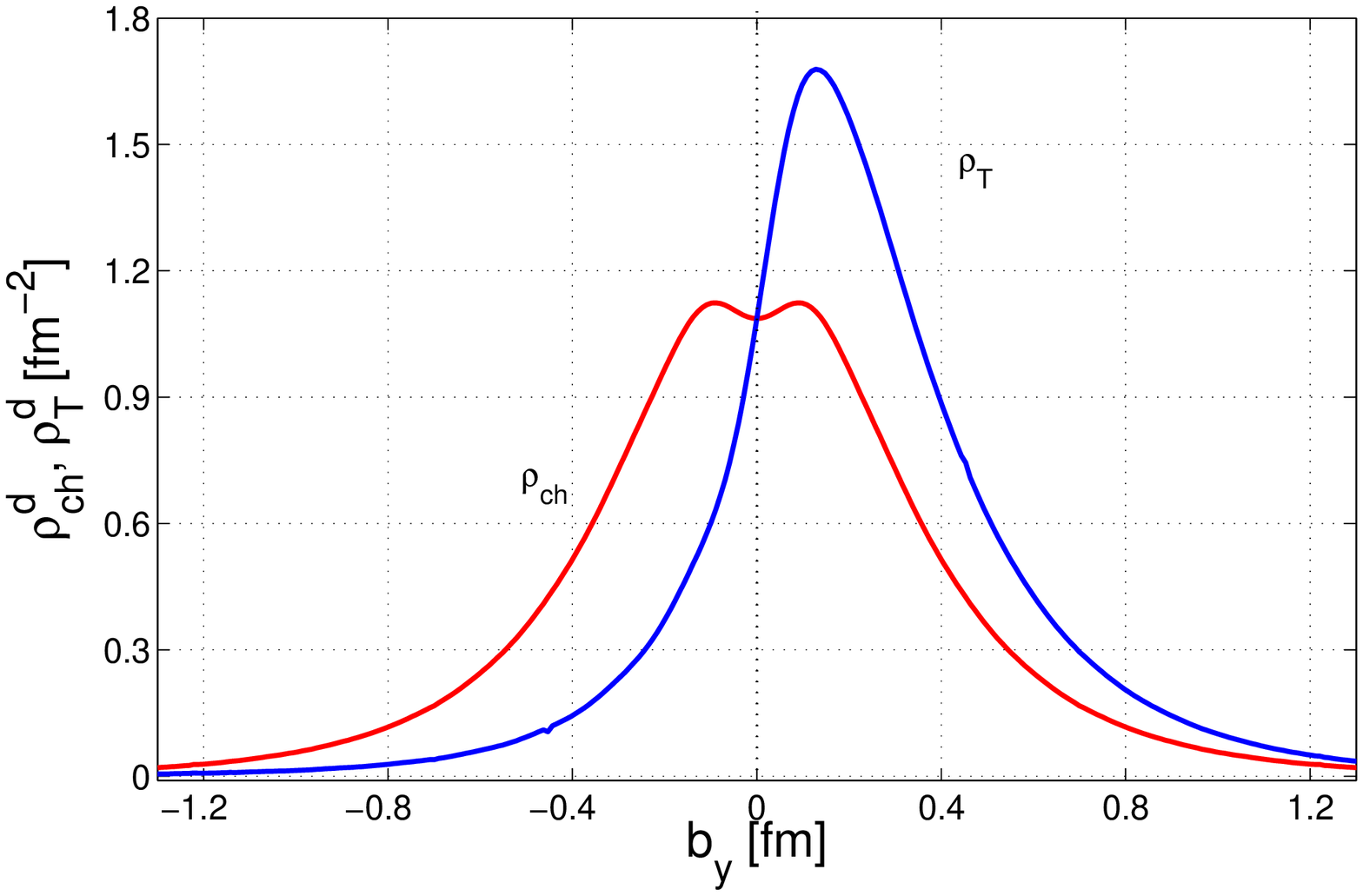}
\end{minipage}
\caption{\label{flavor_TADS} Comparison of the charge densities between the unpolarized and the transversely polarized (a) proton, (b) neutron, (c) up, and (d) down quark charge densities for the unpolarized and the transversely polarized nucleon calculated in the hard-wall AdS/QCD model.}
\end{figure*} 
%%%%%%%%%%%%%%%

We show the charge and anomalous magnetization densities for the unpolarized proton in Figs. \ref{proton_dencities}(a) and \ref{proton_dencities}(b).  
The contributions of the flavors to the proton densities $e_{u/d}\rho_{fch}^{u/d}$ and $e_{u/d}{\rho}_{fm}^{u/d}$ are shown in Figs.  \ref{proton_dencities}(c) and \ref{proton_dencities}(d). 
Similarly, the charge and anomalous magnetization densities for the unpolarized neutron are shown in Figs. \ref{neutron_densities}(a) and \ref{neutron_densities}(b), and the corresponding contributions of flavors $e_{d/u}\rho_{fch}^{u/d} $ and $e_{d/u}{\rho}_{fm}^{u/d}$ are shown in Figs. \ref{neutron_densities}(c) and \ref{neutron_densities}(d). The plots suggest that the predictions of the hard-wall AdS/QCD model for the unpolarized transverse densities are more or less in agreement with the two different global parametrizations Kelly \cite{kelly04} and Bradford $et$ $al.$ \cite{brad}. At the center of mass $(b=0)$, the hard-wall AdS/QCD prediction shows a little higher value of the charge densities compared to the parametrizations. We should mention here that the soft-wall AdS/QCD models \cite{AC,BT2} fail to reproduce the neutron charge density at small $b$ as shown in \cite{CM3}.
The unpolarized charge density for the neutron [Fig. \ref{neutron_densities}(a)] displays a behavior having  a negatively charged core surrounded by a  ring of positive charge density, and the negative interior core shifts towards the center of mass in the case of the hard-wall AdS/QCD model. 
The contribution of the $u$ quark in the proton charge density is large enough compared to the $d$ quark, whereas the contribution of the $u$ quark is almost twice that of the $d$ quark in the neutron charge density. In the case of anomalous magnetization density, the $d$ quark contribution in the neutron is quite high compared to the $u$ quark. 
The charge and anomalous magnetization densities of the individual quarks for the unpolarized nucleon are shown in Fig. \ref{flavors}. 
Again at small $b$, the hard-wall AdS/QCD model disagrees with the parametrizations of Kelly \cite{kelly04} and Bradford $et$ $al.$ \cite{brad} for both quark charge densities and gives a higher value. For the quark anomalous magnetization densities, the hard-wall AdS/QCD model is in good agreement with the parametrizations.
%%%%%%%%%%%%%%%%...top view...%%%%%%%%%%%%%%%%%%%%
\begin{figure*}[htbp]
\begin{minipage}[c]{0.98\textwidth}
\small{(a)}
\includegraphics[width=7.3cm,height=5.6cm,clip]{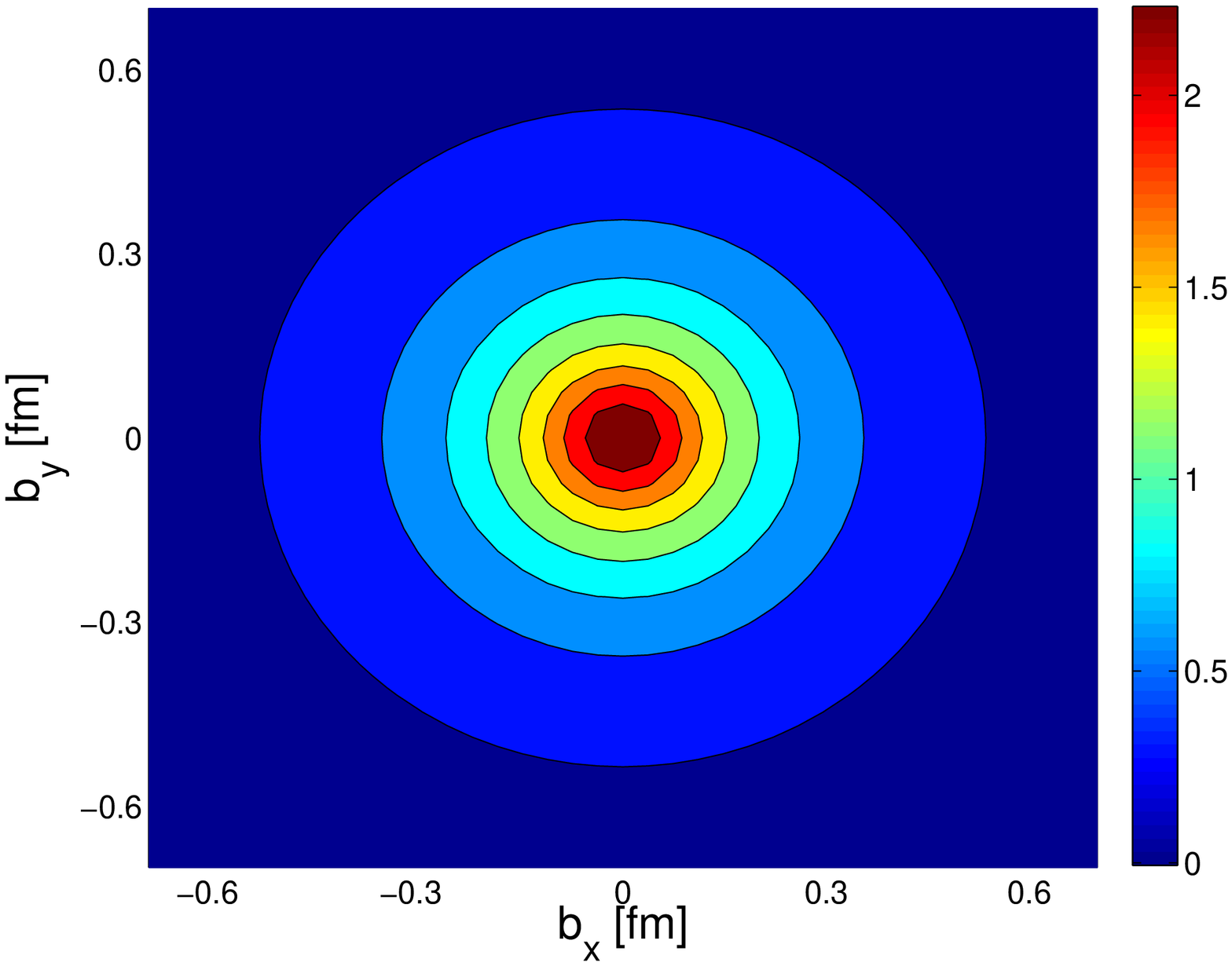}
\hspace{0.1cm}%
\small{(b)}\includegraphics[width=7.3cm,height=5.6cm,clip]{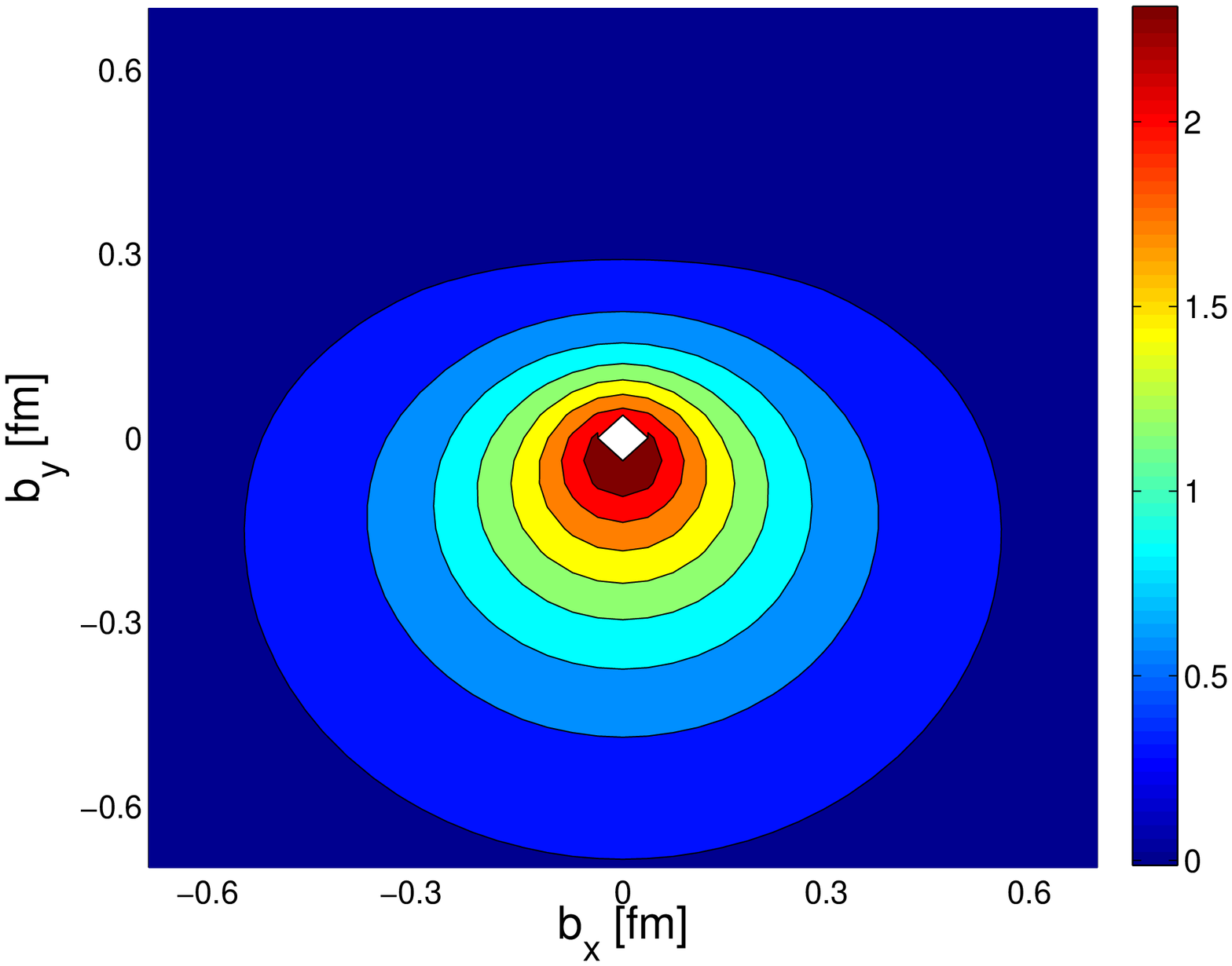}
\end{minipage}
\begin{minipage}[c]{0.98\textwidth}
\small{(c)}
\includegraphics[width=7.3cm,height=5.6cm,clip]{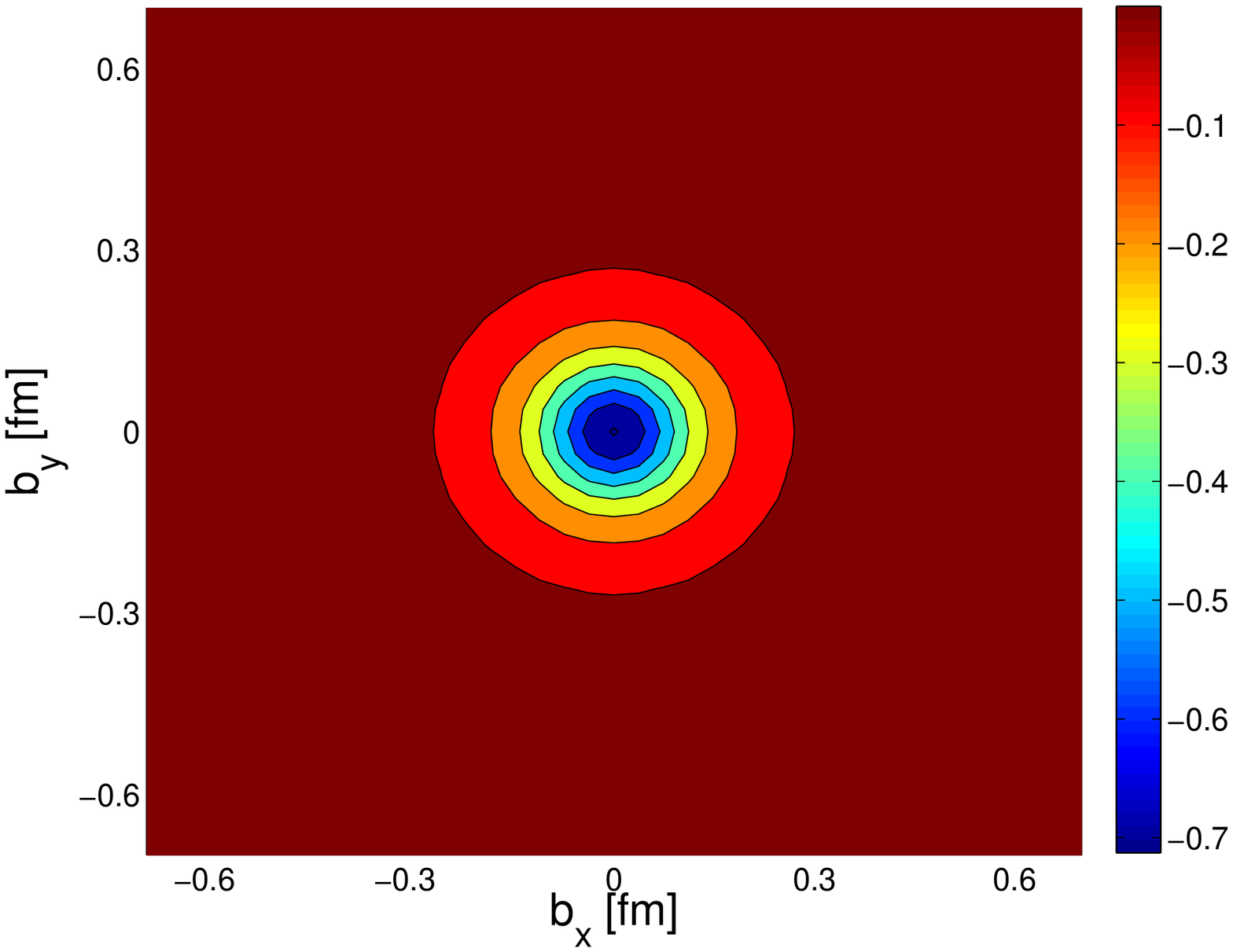}
\hspace{0.1cm}%
\small{(d)}\includegraphics[width=7.3cm,height=5.6cm,clip]{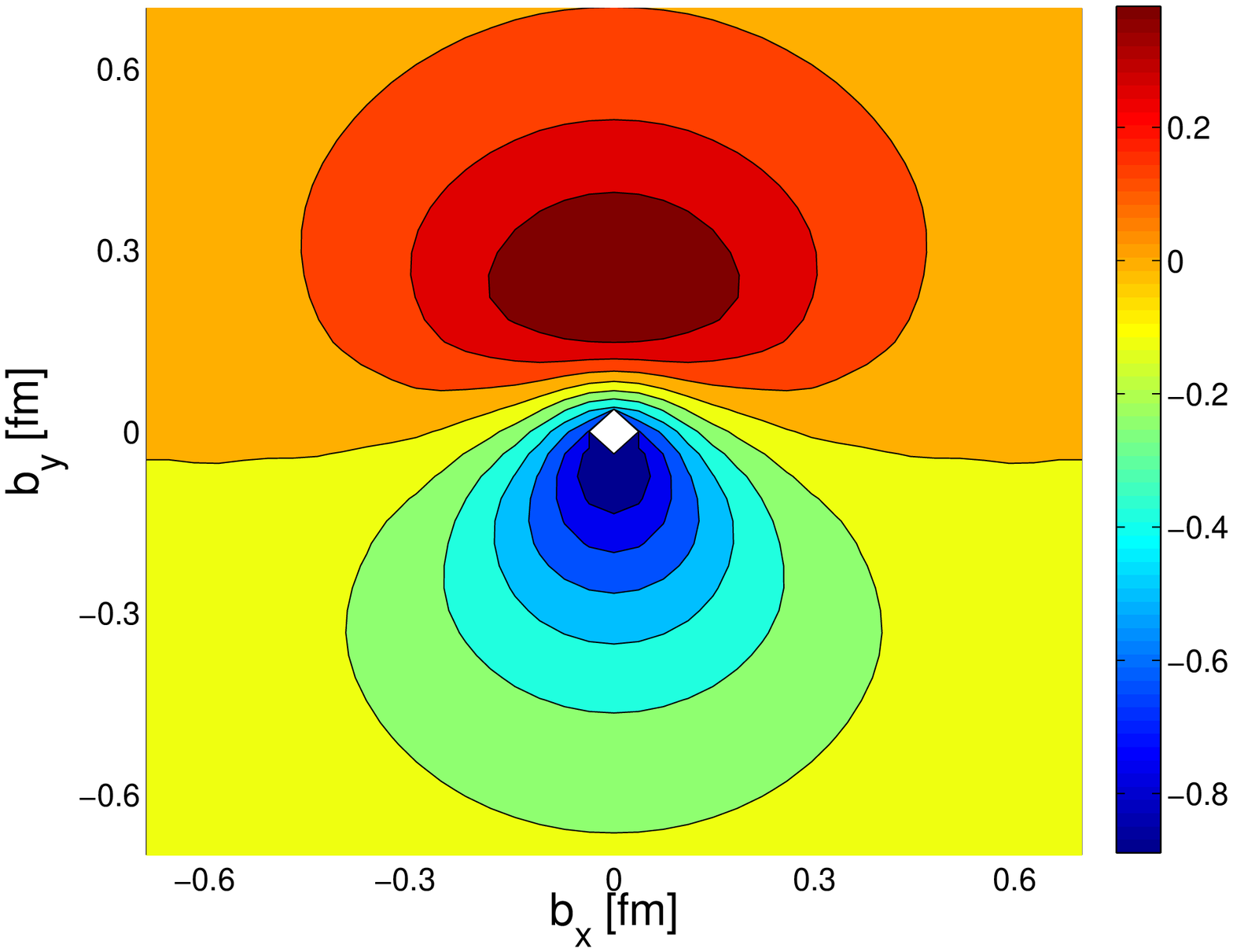}
\end{minipage}
\caption{\label{nucleons_top} The charge  densities in the transverse plane for  the (a) unpolarized  proton, (b) transversely polarized  proton, (c) unpolarized neutron, and (d) transversely polarized neutron. Transverse polarization is along the $x$ direction. }
\end{figure*} 
%%%%%%%%%%%%%%%%%%%%%%%%%%...flavors..%%%%%%%%%%%%
\begin{figure*}[htbp]
\begin{minipage}[c]{0.98\textwidth}
\small{(a)}
\includegraphics[width=7.3cm,height=5.6cm,clip]{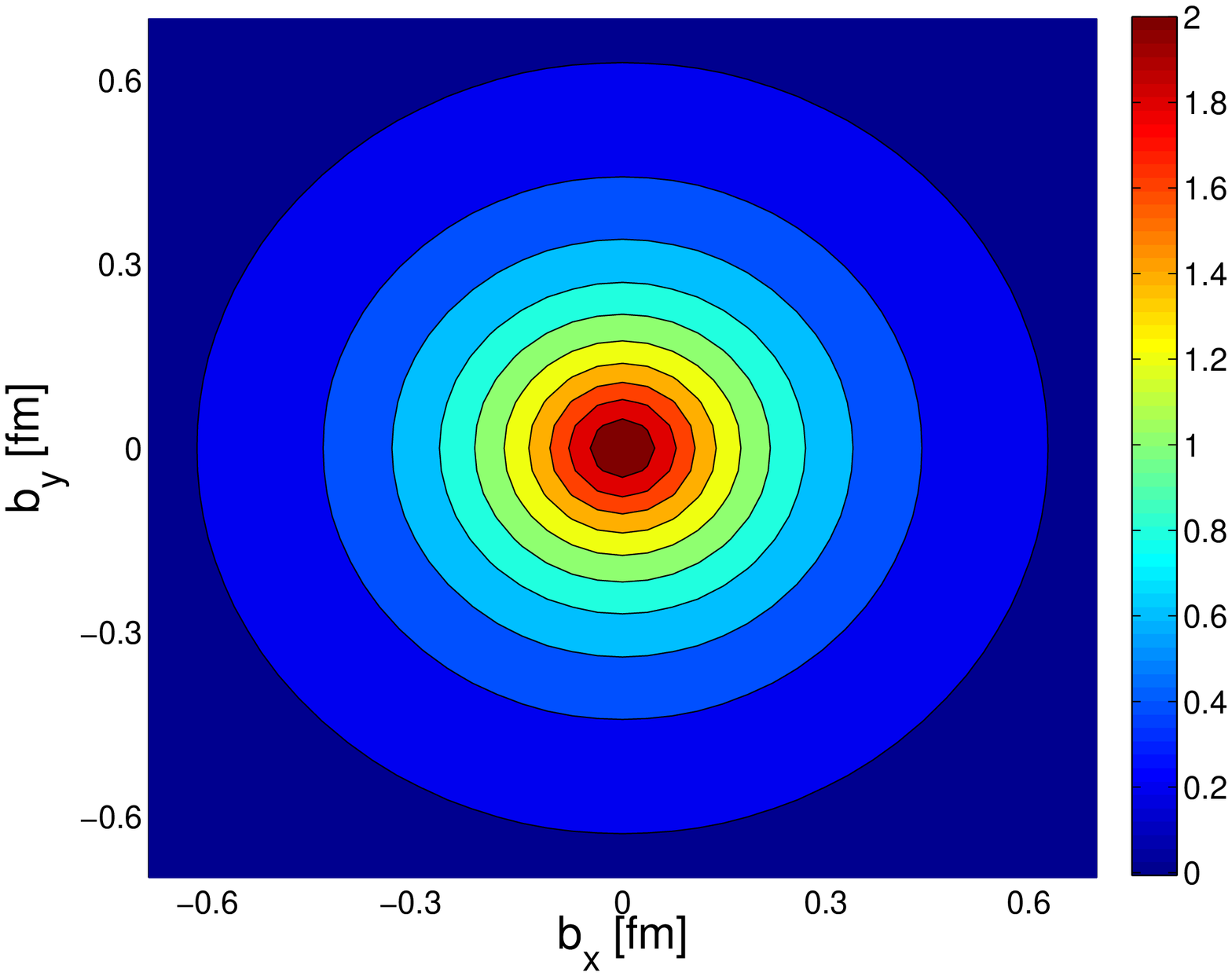}
\hspace{0.1cm}%
\small{(b)}\includegraphics[width=7.3cm,height=5.6cm,clip]{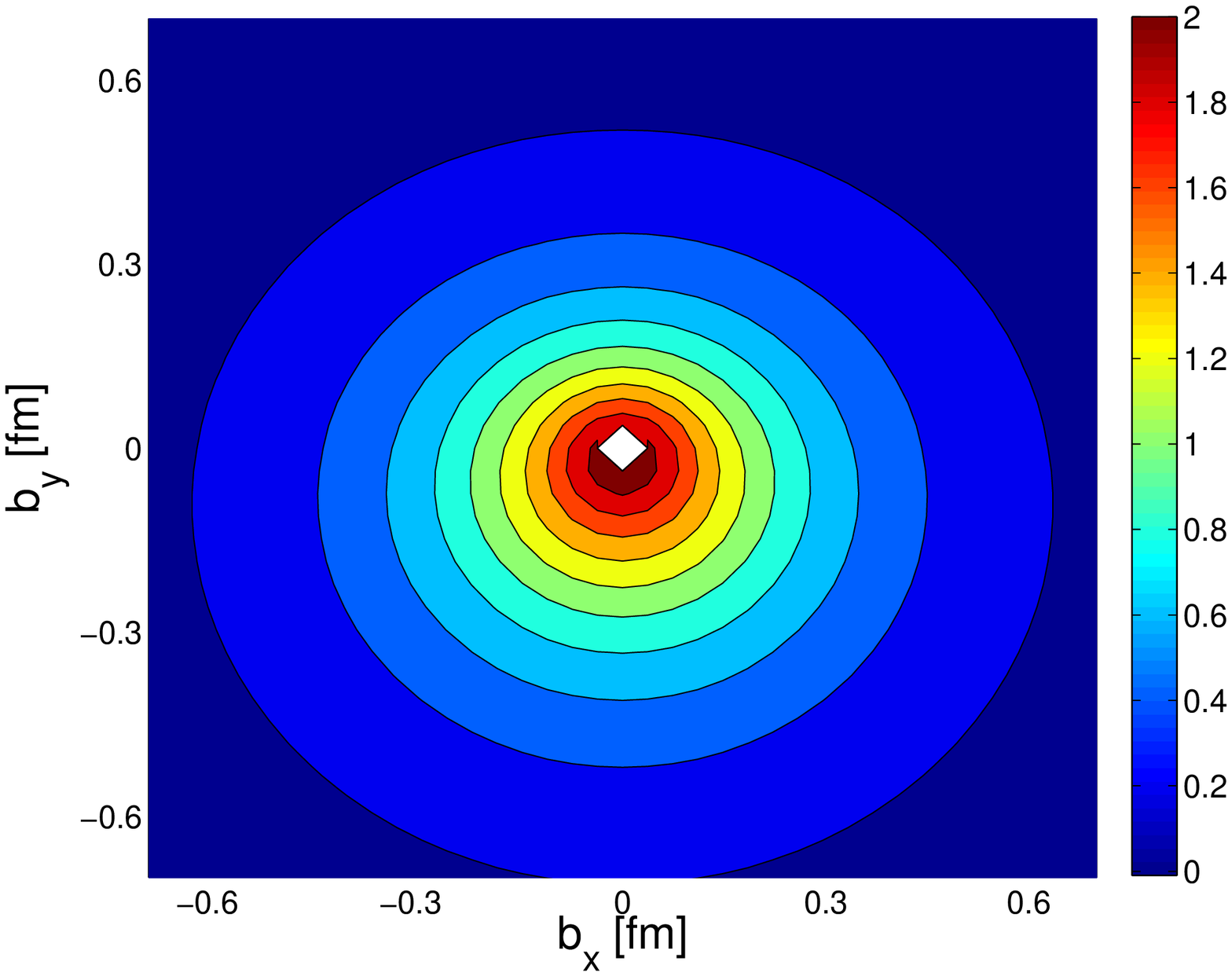}
\end{minipage}
\begin{minipage}[c]{0.98\textwidth}
\small{(c)}
\includegraphics[width=7.3cm,height=5.6cm,clip]{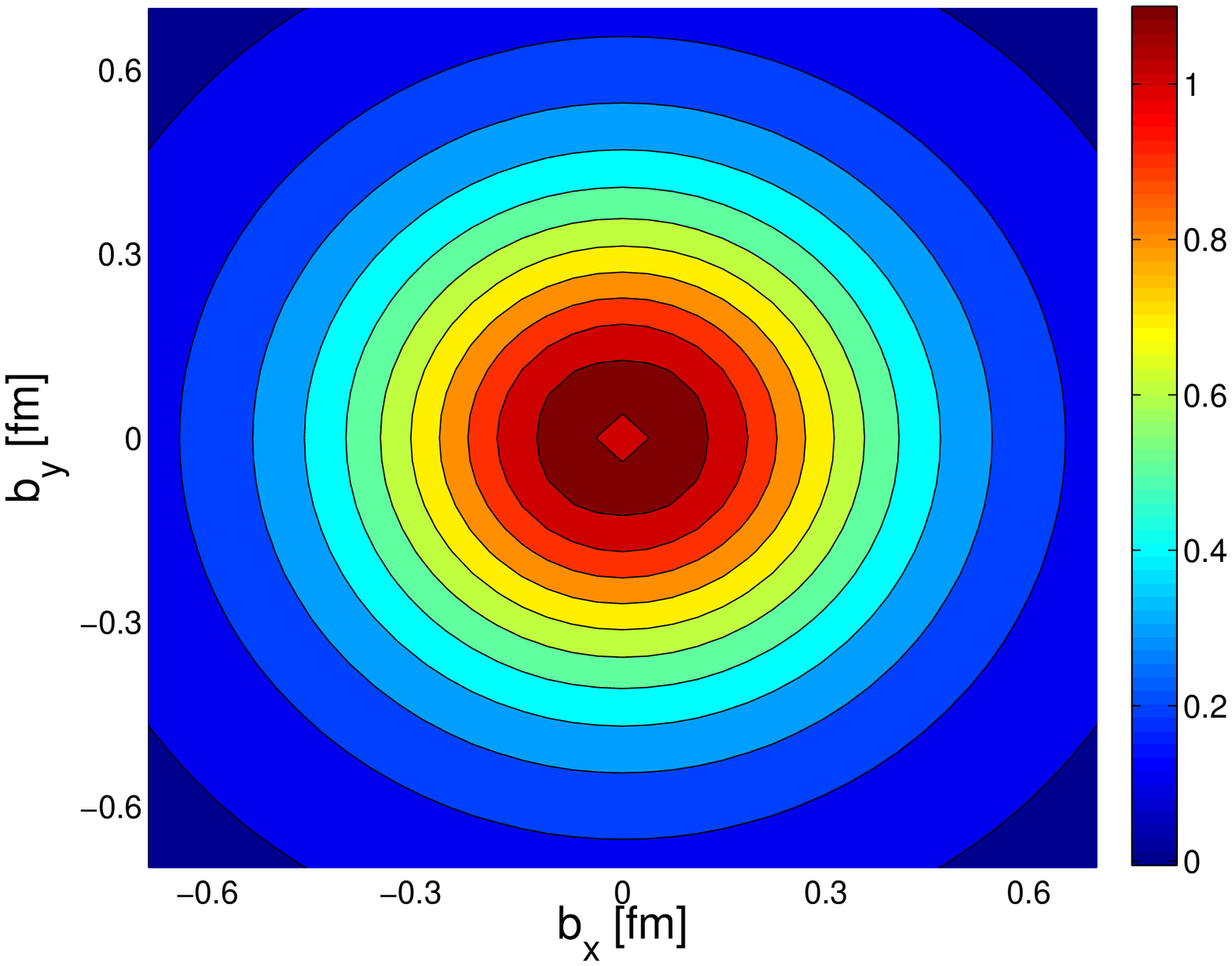}
\hspace{0.1cm}%
\small{(d)}\includegraphics[width=7.3cm,height=5.6cm,clip]{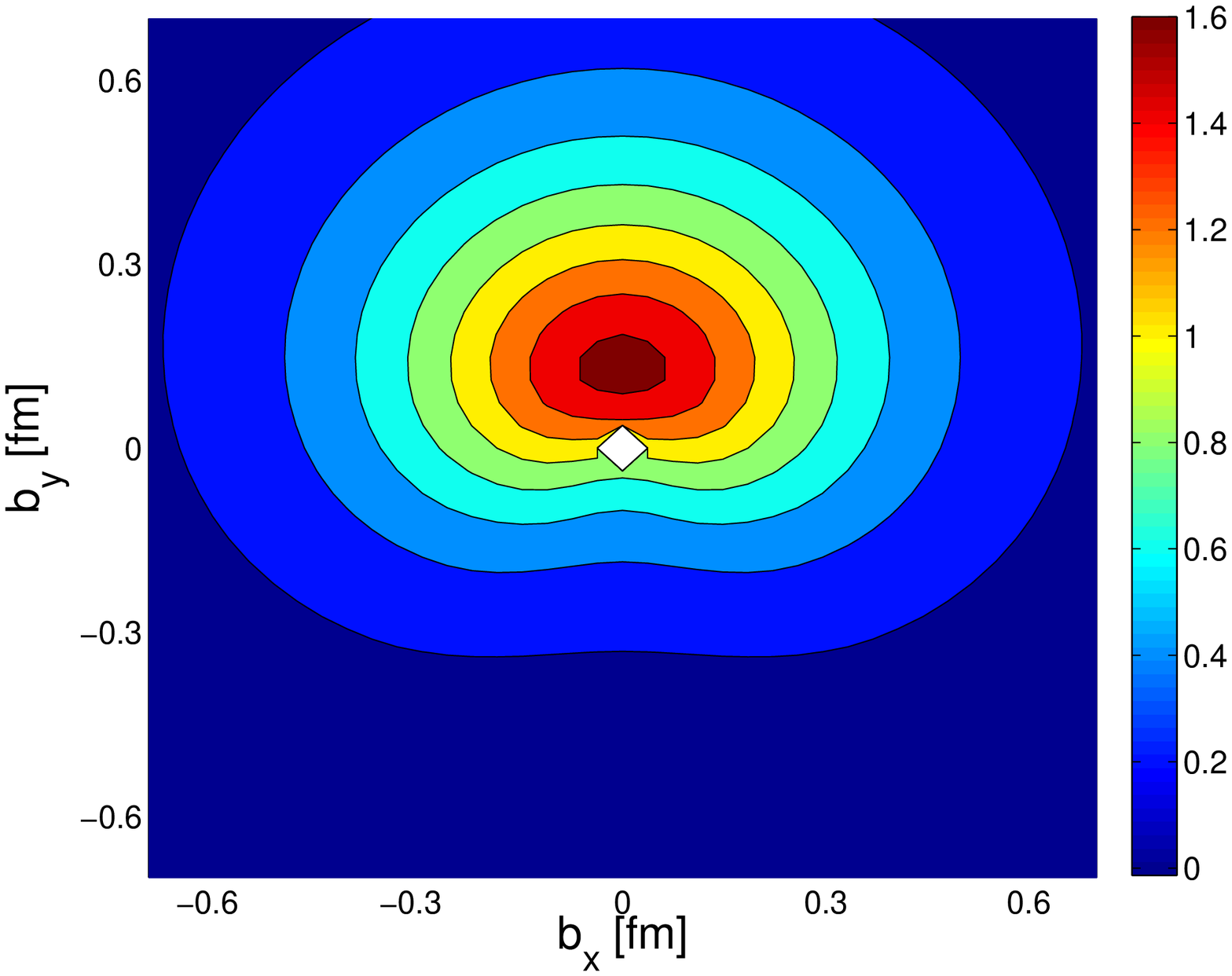}
\end{minipage}
\caption{\label{flavor_top} The charge  densities in the transverse plane of the $u$ quark for (a) unpolarized nucleon and (b) transversely polarized nucleon, and the $d$ quark for (c) unpolarized nucleon and (d) transversely polarized nucleon. Polarization is along the $x$ direction.}
\end{figure*} 
%%%%%%%%%%%%%%%%%%%%%%%%%%%%%%%%%%%%%%%%%%%%%%%%%%
%%%%%%%%%%%%%%%%%%%%%%%%%%%%%%
\subsection{Charge densities for transversely polarized nucleon}
%%%%%%%%%%%%%%%%%%%%%%%%%%%%%%
The charge density in the transverse plane for a transversely polarized nucleon is given by\cite{vande,CM3}
\be
\rho_T(b)
=\rho_{ch}-\sin(\phi_b-\phi_s)\frac{1}{2M b}\rho_m\label{trans_pol},
\ee
where the mass of nucleon is $M$, the transverse polarization direction of the nucleon is denoted by
$S_\perp=(\cos\phi_s \hat{x}+\sin\phi_s\hat{y})$, and the transverse impact parameter is given by $b_\perp=b(\cos\phi_b \hat{x} +\sin\phi_b\hat{y})$.  We take the polarization direction of the nucleon along the $x$ axis, ie., $\phi_s=0$.  
The first term in Eq.(\ref{trans_pol}) is the unpolarized charge density and the second term gives an indication of the deviation from the circular symmetry of the unpolarized charge density \cite{vande}.
We show the charge densities for the proton and neutron polarized along the $x$ axis in Figs. \ref{flavor_T}(a) and \ref{flavor_T}(b). The individual $u$ and $d$ quark charge  densities for the nucleon polarized along the $x$ axis are shown in Figs. \ref{flavor_T}(c) and \ref{flavor_T}(d). Again, the AdS/QCD model is in good agreement with the parametrizations, except that it provides larger peak densities than the parametrizations. 

We compare the AdS/QCD results of the charge densities for the unpolarized  and transversely polarized  proton in Fig. \ref{flavor_TADS}(a) and a similar plot for the neutron is shown in Fig. \ref{flavor_TADS}(b). 
As expected, the deviation of the transversely polarized density from the unpolarized density is quite large for the neutron compared to the proton. This is because of much higher anomalous magnetization density than the unpolarized charge densities for the neutron, whereas for the proton, the unpolarized charge density large enough compared to the anomalous magnetization density. Similar behavior has been seen in \cite{vande,miller10,silva,CM3}. 
The $u$ and $d$ quark charge densities for the transversely polarized and unpolarized nucleon are shown
in Figs. \ref{flavor_TADS}(c) and \ref{flavor_TADS}(d). Because of a similar reason, as stated before for nucleons, the deviation from the symmetric unpolarized density is more for the $d$ quark than for the $u$ quark. One can notice that the shifting of the charge density is the opposite for the $u$ and $d$ quarks. The reason is that the anomalous magnetization density is negative for the $d$ quark but positive for the $u$ quark. A top view of three-dimensional charge densities for nucleons in the transverse plane is shown in Fig. \ref{nucleons_top}.  
Figs. \ref{nucleons_top}(a) and \ref{nucleons_top}(b) represent the charge densities for the unpolarized proton and proton polarized along the $x$ direction. Figs. \ref{nucleons_top}(c) and \ref{nucleons_top}(d) represent the same for the neutron. The unpolarized charge densities are axially symmetric. One notices that the charge density for the transversely polarized proton gets displaced towards the negative $b_y$ direction, and due to  large negative anomalous magnetic moment, which leads to an induced electric dipole moment in the $y$-direction, the charge density for the transversely polarized neutron shows a dipole pattern.
In Figs. \ref{flavor_top}(a)-\ref{flavor_top}(d), we show the top view of the  $u$ and $d$ quarks three-dimensional charge densities in the transverse plane for both the unpolarized and transversely polarized nucleons. It shows that the displacement of the charge density is more in the case of the $d$ quark and opposite the direction of the $u$ quark.

%%%%%%%%%%%%%%%%%%%%%%%%%%%%%%%%%%%%%%%%%%%%%%%%%%%%%%%%%%%%%%%%%%%%%%%%%%%%
\vskip0.2in
\noindent
\section{\bf Summary}\label{concl}
In this paper, we have studied the flavor decompositions of the nucleon form factors for the $u$ and $d$ quarks in a hard-wall AdS/QCD  model, and
the consequences are compared with the experimental data and with two different soft-wall AdS/QCD models.
It has been observed that the Dirac and Pauli form factors of each flavor in this model deviate at higher $Q^2$ from the experimental data. Compared with the soft-wall models, it can be concluded that for $F_1^d$ only, the hard-wall AdS/QCD model gives a better description than the soft I. However, the overall description of $F_1^d$ predicted by the soft II is better, particularly at higher $Q^2$. For $F_1^u$ and $F_2^{u/d}$, soft I describes the data much better compared to the hard-wall and the other soft-wall models. Again, the ratios of the flavor form factors such as $F_1^d/F_1^u$ and $G_E^d/G_M^d$ in the hard-wall model agree well with the experimental data but the ratios which involve $F_1^d$ are not well described by the soft-wall models. It can also be noted that in the higher $Q^2$ region, the hard-wall model generates better data of $Q^2F_2^p/F_1^p$ as compared to the soft-wall models, whereas at low $Q^2$, the result is better in the soft-wall models than in the hard-wall model.    

We have also presented a detailed study of the transverse charge and anomalous magnetization densities for the nucleons as well as the flavor decompositions of the densities in the same hard-wall AdS/QCD  model. The results have been compared with the two standard phenomenological parametrizations  of the form factors. We have considered both the unpolarized and the transversely polarized nucleons in this work. Our analysis shows that the AdS/QCD model is in good agreement with the parametrizations at higher $b$ but deviates at lower $b$. The unpolarized densities are axially symmetric in the transverse plane, while the densities for the transversely polarized nucleons get displaced along the $y$ direction if the nucleon is polarized along the $x$ direction. This hard-wall AdS/QCD  model produces much 
better result for the  charge density of the unpolarized neutron than the soft-wall AdS/QCD models as shown in \cite{CM3}.       
The charge density for the transversely polarized neutron shows a dipole pattern in the transverse plane. We have also studied the transverse charge and anomalous magnetization densities for individual  $u$ and $d$ quarks. It has been observed that the distortion in the $d$ quark charge density is much stronger than that for the $u$ quark, and the densities get shifted in the opposite direction of each other for the transversely polarized nucleon.
%%%%%%%%%%%%%%%%%%%%%%%%%%%
%\textbf{\textit{Acknowledgements:}}
\acknowledgments
%%%%%%%%%%%%%%%%%%%%%%%%%%%
The author thanks Dipankar Chakrabarti for many useful discussions. %and giving valuable suggestions.
%%%%%%%%%%%%%%%%%%%%%%%%%%%%%%%%%%%%%%%%%%%%%%%%%%%%%%%%%%%%%%%%%%%%%%%

%%%%%%%%%%%%%%%%%%%%%%%%%%%%%%%%%%%%%%%%%%%%%%%%%%%%%%%%%%%%%%%%%%%%%%%%%%%%%%%%%%

%%%%%%%%%%%%%%%%%%%%%%%%%%%%%
%%%%%%%%%%%%%%%%%%%%%%%%%%%%%%%%%%

%%%%%%%%%%%%%%
\end{document}